\documentclass[10pt,twocolumn,letterpaper]{article}

\usepackage{cvpr}              

\usepackage[dvipsnames]{xcolor}

\definecolor{cvprblue}{rgb}{0.21,0.49,0.74}

\usepackage[pagebackref,breaklinks,colorlinks,citecolor=cvprblue]{hyperref}

\usepackage{float}
\usepackage{multirow}
\usepackage{graphicx}
\usepackage{colortbl}
\usepackage{utfsym}
\usepackage{subcaption}
\usepackage{tabularx}
\usepackage{booktabs}

\usepackage{amsmath}
\usepackage{amssymb}
\usepackage{makecell}

\def\paperID{4359} 
\def\confName{CVPR}
\def\confYear{2024}

\definecolor{r1}{RGB}{0, 128, 0} 
\definecolor{r2}{RGB}{34,150,255} 
\definecolor{r3}{RGB}{255, 165, 0} 
\definecolor{t}{RGB}{206,70,118} %
\definecolor{red}{RGB}{255, 0, 0} 

\definecolor{color3}{rgb}{0.95,0.95,0.95}

\title{Low-Res Leads the Way: \\ Improving Generalization for Super-Resolution by Self-Supervised Learning}

\author{
Haoyu Chen$^{1}$,\quad Wenbo Li$^{2}$, \quad Jinjin Gu$^{3}$,\quad Jingjing Ren$^{1}$,\quad Haoze Sun$^{4}$,\\ \vspace{1.2mm} Xueyi Zou$^{2}$,\quad Zhensong Zhang$^{2}$,\quad Youliang Yan$^{2}$,\quad Lei Zhu$^{1,5}$\thanks{Lei Zhu (leizhu@ust.hk) is the corresponding author.}\\ \vspace{-0.5mm}
\footnotesize $^{1}$The Hong Kong University of Science and Technology (Guangzhou)\quad
\footnotesize $^{2}$Huawei Noah’s Ark Lab\quad $^{3}$The University of Sydney\\ \vspace{-0.5mm}
\footnotesize $^{4}$Tsinghua University\quad
\footnotesize $^{5}$The Hong Kong University of Science and Technology\\
{\tt\small Project page: \url{https://haoyuchen.com/LWay}}}

\begin{document}

\maketitle

\begin{abstract}
\vspace{-3mm}
For image super-resolution (SR), bridging the gap between the performance on synthetic datasets and real-world degradation scenarios remains a challenge. This work introduces a novel "Low-Res Leads the Way" (LWay) training framework, merging Supervised Pre-training with Self-supervised Learning to enhance the adaptability of SR models to real-world images. Our approach utilizes a low-resolution (LR) reconstruction network to extract degradation embeddings from LR images, merging them with super-resolved outputs for LR reconstruction. Leveraging unseen LR images for self-supervised learning guides the model to adapt its modeling space to the target domain, facilitating fine-tuning of SR models without requiring paired high-resolution (HR) images. The integration of Discrete Wavelet Transform (DWT) further refines the focus on high-frequency details. Extensive evaluations show that our method significantly improves the generalization and detail restoration capabilities of SR models on unseen real-world datasets, outperforming existing methods. Our training regime is universally compatible, requiring no network architecture modifications, making it a practical solution for real-world SR applications.

\vspace{-5mm}
\end{abstract}    
\vspace{-2mm}
\section{Introduction}
\vspace{-2mm}
\label{sec:intro}

\begin{figure}[!tbp]
  \centering
  \includegraphics[width=1\linewidth]{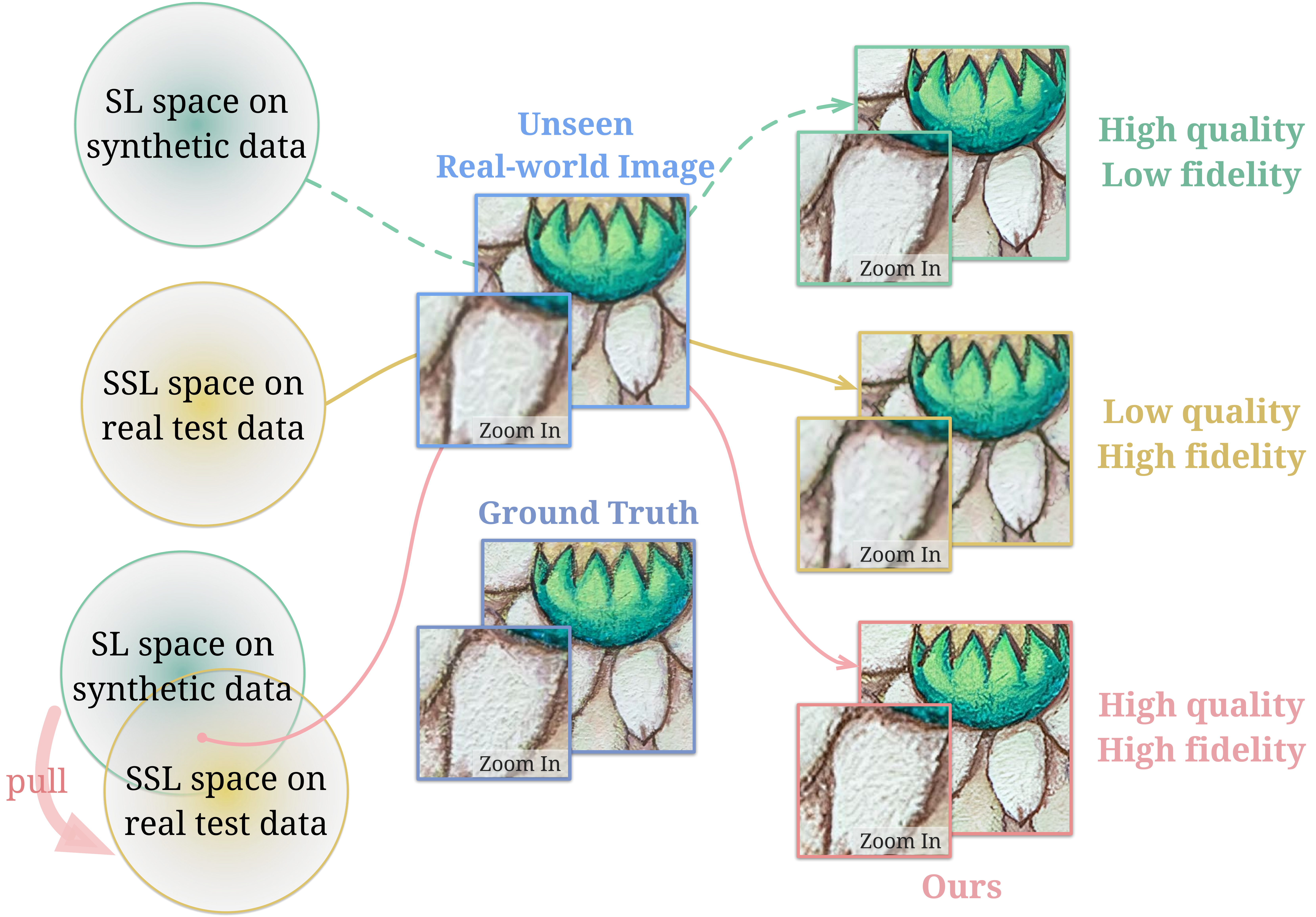} 
  \vspace{-7mm}
  \caption{Our proposed training method combine the benefits of supervised learning (SL) on synthetic data and self-supervised learning (SSL) on the unseen test images, achieve high quality and high fidelity SR results.}
  \label{fig:motivation}
\vspace{-6mm}
\end{figure}

Image super-resolution (SR) aims to restore high-resolution (HR) images from their low-resolution (LR) or degraded counterparts. The inception of the deep-learning-based SR model can be traced back to SRCNN~\cite{dong2015image}. Recently, advancements in deep learning models have substantially enhanced SR performance~\cite{liang2021swinir,ijcai2023p121,chen2023activating,chen2021attention,zhang2018residual, ahn2018fast,dai2019second,zhang2019residual, sun2023coser,Zou_2022_CVPR,chen2023dual,chen2023recursive,li2022blueprint,zhou2022efficient}, particularly in addressing specific degradation types like bicubic downsampling. Nevertheless, the efficacy of SR models is generally restricted by the degradation strategies employed during the training phase, posing great challenges in complex real-world applications.

\begin{figure}[!tbp]
  \centering
  \includegraphics[width=1\linewidth]{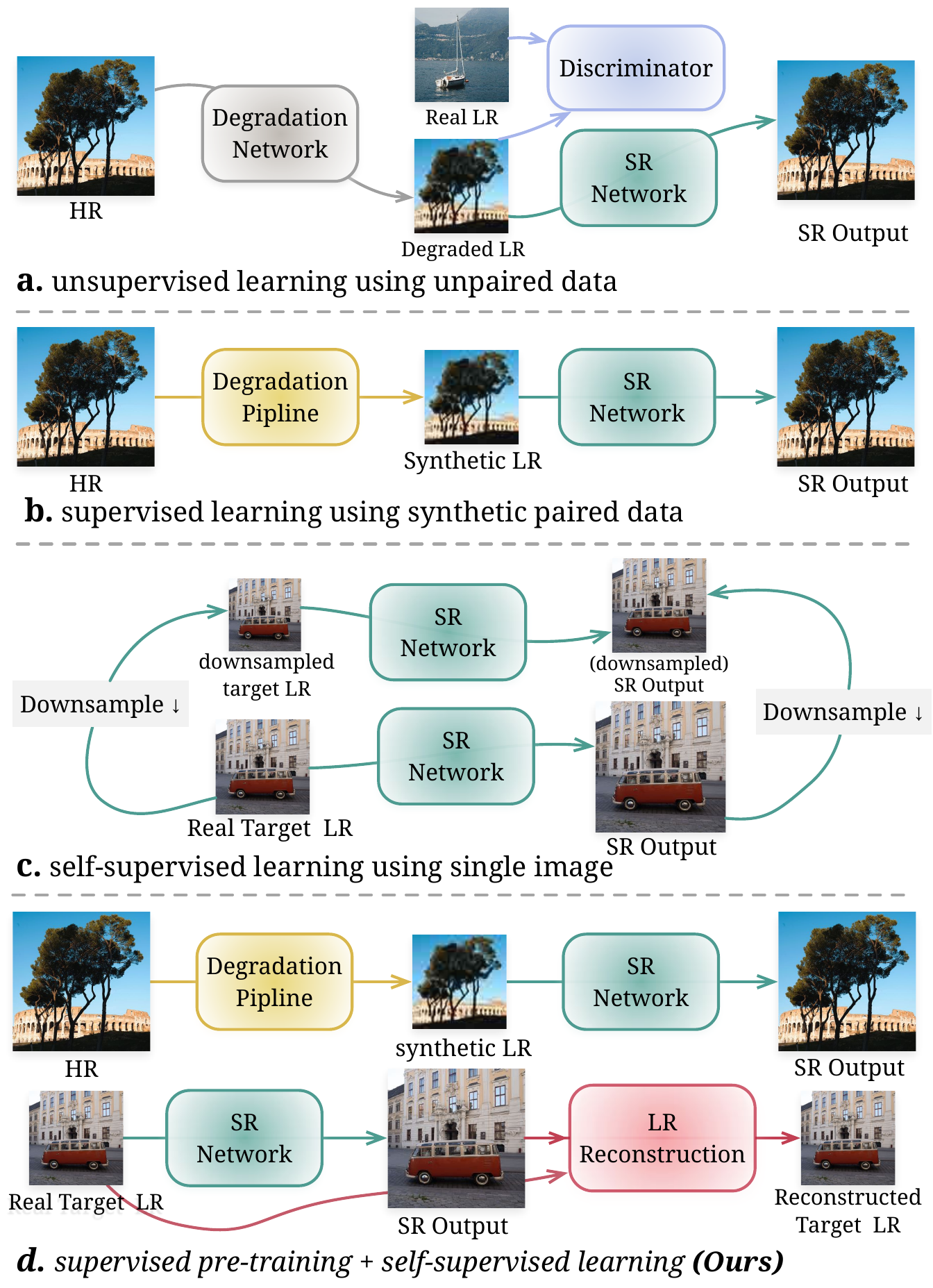} 
  \vspace{-7mm}
  \caption{Comparison of different learning approaches for real-world image SR.}
  \vspace{-5mm}
  \label{fig:comparison}
\end{figure}

In the realm of real-world SR, as shown in \figurename~\ref{fig:comparison}, training approaches can primarily be categorized into three main paradigms.
\textbf{(a) Unsupervised Learning with Unpaired Data}: Methods within this paradigm~\cite{ulyanov2018deep, yuan2018unsupervised, zhu2017unpaired, bell2019blind, soh2020meta, wei2021unsupervised, fritsche2019frequency, bulat2018learn} commonly utilize Generative Adversarial Networks (GAN) architecture to learn target distributions without paired data. Using one or multiple discriminators, they distinguish between generated images and actual samples, guiding the generator to model accurately. 
However, as this approach heavily relies on external data, it encounters significant challenges when facing scarce target domain data, particularly in real-world scenarios. 
The GAN framework for unsupervised learning also has some drawbacks. Firstly, it inherently struggles with stability during training, leading to noticeable artifacts in SR outputs. 
Secondly, it is difficult for a single 0/1 plane modelled by a discriminator to accurately separate the target domain~\cite{liu2022blind}. This can result in imprecise distribution learning.
\textbf{(b) Supervised Learning with Paired Synthetic Data}: BSRGAN~\cite{zhang2021designing} and Real-ESRGAN~\cite{wang2021real} have largely enhanced the SR model's generalization ability by simulating more realistic degradation. However, synthetic data, despite mimicking certain real-world conditions, inadequately captures the complex and variable nature of real scenarios, the gap between synthetic and real degradation persists. Consequently, the limited degradation patterns in synthetic data may lead to an over-smoothness issue, sacrificing crucial details and textures. Adapting effectively to complex, variable, or unknown degradations thus remains a formidable challenge.
\textbf{(c) Self-supervised Learning with a Single Image}: Techniques falling within this category~\cite{shocher2018zero, neshatavar2023icf, cheng2020zero} leverage the intrinsic statistical characteristics of natural images, eliminating the necessity for external datasets. Generally, these methods enable self-supervised learning directly from the input LR image. Despite its inherent flexibility, this approach may exhibit reduced efficacy when handling images lacking repetitive patterns. As a result, in real-world scenarios, where necessary recurring structure are absent, these techniques tends to underperform compared to supervised learning methods that employ paired synthetic data.

It's notable that real LR/HR image pairs in the target domain are often prohibitively expensive or unavailable. Furthermore, a significant gap persists between synthesized data and real-world data. Given the intrinsic limitations of current methodologies, a critical question arises: Is there an approach that combines the strengths of these diverse strategies? In addressing this, we propose the novel "Low-Res Leads the Way" (LWay) training framework, which merges supervised learning (SL) pre-training with self-supervised learning (SSL) (see \figurename~\ref{fig:comparison} (d)). This approach aims to narrow the disparity between synthetic training data and real test images, as depicted in \figurename~\ref{fig:motivation}. By integrating supervised learning's predictive capabilities with the ability to swiftly adapt to unique characteristics present in test LR images, this framework effectively produces high-quality results for unseen real-world images.

The initial step involves training an LR reconstruction network specifically designed to extract a degradation embedding from the LR image. This degradation embedding is then applied to the HR image, facilitating the re-generation of LR content. Upon encountering a test image, we derive its super-resolved result from an off-the-shelf SR model pre-trained on synthetic data. This output is fed into the fixed LR reconstruction network to produce the corresponding degraded counterpart. Subsequently, a self-supervised loss is computed by comparing this degraded counterpart to the original LR image, thereby updating specific parameters within the SR model. Given our observation that pre-trained SR models adeptly handle low-frequency domains but falter in high-frequency areas, we incorporate Discrete Wavelet Transform (DWT) to isolate high-frequency elements from the LR image. This component effectively shifts the model's focus to the recuperation of high-frequency nuances, and avoids negative impacts on low-frequency areas.

With this innovative framework, our approach eliminates the need for paired LR/HR target domain images, significantly enhancing the performance of SL pre-trained models on unseen real-world data. Our method not only retains the essential content of LR images but also adds high-definition characteristics, ensuring a balance between fidelity and quality. Moreover, this training regime requires no modifications to the network architecture, offering broad compatibility across all SR models. Through extensive evaluations on real-world datasets, we have demonstrated our method's substantial improvements in generalization performance.

\vspace{-1mm}
\section{Related Work}
\label{sec:relatedwork}

\subsection{Supervised Learning for Real-World SR}
\vspace{-2mm}

While recent years have witnessed significant advancements in the field of super-resolution (SR), conventional SR models such as SRCNN~\cite{dong2015image}, VDSR~\cite{kim2016accurate}, EDSR~\cite{lim2017enhanced}, RCAN~\cite{zhang2018image}, among others~\cite{zhang2018residual, ahn2018fast, dai2019second, chen2021attention, zhang2019residual, li2020lapar, li2022best, liang2021swinir, liu2020residual, kim2016deeply, ijcai2023p121,chen2023dual,chen2023recursive,li2022blueprint,zhou2022efficient}, have predominantly relied upon predefined degradation processes, such as bicubic downsampling. This simplification, while contributing to the theoretical understanding of SR, often falls short in capturing the intricate and diverse degradations inherent in real-world imaging scenarios, limiting practical adaptability across applications. Consequently, there is a pressing need to explore more sophisticated and realistic degradation models.

To this end, recent efforts have been directed toward methods capturing paired low-resolution (LR) and high-resolution (HR) images from real-world environments, as demonstrated by datasets like RealSR~\cite{cai2019toward} and DRealSR~\cite{wei2020component}. However, these methods face challenges, including precise image alignment, complex hardware setups, and specific degradation characteristics (\textit{e.g.}, Canon 5D3 and Nikon D810 cameras in RealSR), posing obstacles to practicality and scalability. Recent techniques, including Real-ESRGAN~\cite{wang2021real} and BSRGAN~\cite{zhang2021designing}, have attempted to address these shortcomings by synthesizing LR images with more realistic degradation. Despite these advancements, a notable disparity persists between synthesized and authentic degradation. This often results in over-smoothed images that sacrifice fine textural details, as illustrated by~\cite{10.5555/3618408.3620129}. Certain studies~\cite{chen2023masked} have endeavored to enhance the generalizability using limited degradation data; however, the practical application scenarios remain restricted.

As a result, there is a growing demand for innovative approaches that are capable of adapting to the intricate and mixed degradation patterns that typify real-world applications. The SR results should not only exhibit high resolution but also encompass rich detail, ensuring fidelity.

\begin{figure*}[!tbp]
  \centering
  \includegraphics[width=1\linewidth]{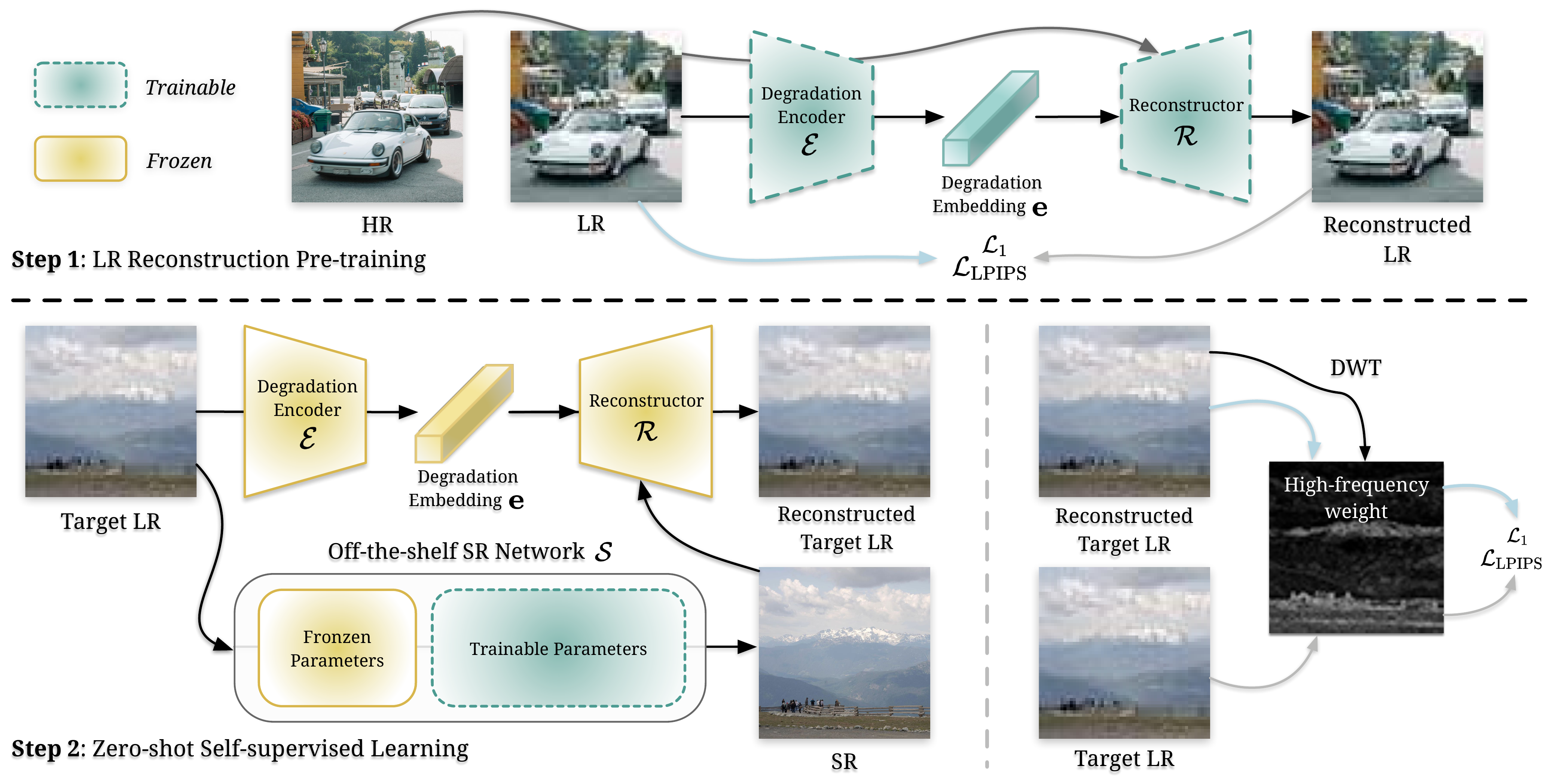} 
  \vspace{-7mm}
  \caption{
  The proposed training pipeline (LWay) consists of two steps. In Step 1, we pre-train a LR reconstruction network to capture degradation embedding from LR images. This embedding is then applied to HR images, regenerating LR content. Moving to Step 2, for test images, a pre-trained SR model generates SR outputs, which are then degraded by the fixed LR reconstruction network. We iteratively update the SR model using a self-supervised learning loss applied to LR images, with a focus on high-frequency details through weighted loss. This refinement process enhances the SR model's generalization performance on previously unseen images.
  }
  \vspace{-4mm}
  \label{fig:method}
\end{figure*}

\vspace{-2mm}
\subsection{Unsupervised Learning for Real-world SR}
\vspace{-2mm}

Unsupervised super-resolution~\cite{ulyanov2018deep, yuan2018unsupervised, zhu2017unpaired, bell2019blind, soh2020meta, wei2021unsupervised, fritsche2019frequency, bulat2018learn} serves as a technique to mitigate generation bias inherent in synthetic datasets. These approaches deviate from the conventional reliance on extensive paired data by harnessing the data-generating capabilities inherent in convolutional neural networks (CNNs). Ulyanov et al.~\cite{ulyanov2018deep} posited CNNs as implicit priors for capturing natural image statistics, a concept further explored by the Zero-Shot Super-Resolution (ZSSR)~\cite{shocher2018zero} model, which uniquely tailors SR algorithms to the repeating patterns within the input image itself. 
Generative Adversarial Networks (GANs) have significantly propelled the field forward. KernelGAN~\cite{bell2019blind}, for instance, aligns the statistical distribution of downscaled images with their original versions, enhancing the refinement of SR methods' outputs. 
CinCGAN~\cite{yuan2018unsupervised} marks an early exploration into utilizing unpaired data for implicit degradation modeling. It employs a strategy that transforms LR images into noise-free `clean' states through bicubic downsampling. This approach, backed by a dual CycleGAN architecture~\cite{zhu2017unpaired}, fosters a cycle-consistent adaptation that eliminates the need for paired datasets. 
%
The unsupervised approach utilizing GANs also encompasses methods such as Degradation GAN~\cite{bulat2018learn}, FSSR~\cite{fritsche2019frequency}, DASR~\cite{wei2021unsupervised} and pseudo-supervision~\cite{maeda2020unpaired}, which all employ discriminators to learn the distributions of HR or LR images, or even clean LR images. These methods are instrumental in constraining the network to transform the generated images to align with the corresponding distributions.

%
Despite considerable advancements in unsupervised methods, they still exhibit certain limitations. 
For instance, ZSSR and similar methods typically rely on the prerequisite assumption that images possess repetitive patterns. 
GAN-based approaches, in particular, require substantial data to fit certain specific degradation types effectively. 
They also face stability challenges during training, which often results in artifacts in SR outputs. Furthermore, the challenge for a discriminator to accurately distinguish the target domain using a binary (0/1) plane model can lead to imprecise learning of distributions.
These constraints pose challenges to the practical utility of these methods in real-world scenarios. Exploring more generalized and flexible approaches becomes imperative.

\section{Method}
\vspace{-1mm}

In the pursuit of practical applications for image SR, we introduce an unprecedented training methodology. This novel strategy marks a departure from established paradigms, fusing the precision of supervised pre-training with the innovation of self-supervised learning to address the complexities of real-world image degradation. 
Our proposed framework is detailed in Figure \ref{fig:method}.

\subsection{LR Reconstruction Pre-training}
\vspace{-1mm}

We introduce an LR reconstruction branch that plays a pivotal role in finetuning our SR model \( \mathcal{S} \) on test images derived from real-world environments.
Central to this process is the Degradation Encoder \( \mathcal{E} \), engineered to distill the degradation signatures from LR images \( I_{\text{LR}} \) into a concise degradation embedding \( \mathbf{e} \). The dimension is 512, formulated as \( \mathbf{e} = \mathcal{E}(I_{\text{LR}}) \). 
Subsequently, the Reconstructor \( \mathcal{R} \) employs \( \mathbf{e} \) and a high-resolution image \( I_{\text{HR}} \) to synthesize an estimated LR image \( \hat{I}_{\text{LR}} \), aiming to fulfill \( \hat{I}_{\text{LR}} = \mathcal{R}(I_{\text{HR}}, \mathbf{e}) \). To ensure the integrity of \( \mathbf{e} \), we incorporate a dual-component loss function \( \mathcal{L} \), integrating both an L1 norm and the Learned Perceptual Image Patch Similarity (LPIPS) metric. The combined loss function is thus articulated as \( \mathcal{L}(I_{\text{LR}}, \hat{I}_{\text{LR}}) = \mathcal{L}_1 + \mathcal{L}_{\text{LPIPS}} \), meticulously tuning the reconstruction fidelity.
Notably, LR reconstruction branch has great robustness, requiring only minimal data for training, is precisely why we advocate for the inclusion of an LR reconstruction branch. This ensures that even when faced with new forms of degradation, its support in the finetuning of the SR model remains uncompromised. The efficiency and robustness of this approach, pivotal in our methodology, will be detailed and validated in the following sections.

\vspace{-1mm}
\subsection{Self-supervised Learning on Test Images}
\vspace{-2mm}
Our approach innovatively fine-tunes a subset of parameters in a SR network, specifically tailored for processing previously unseen real-world images. This method refines the SR network to adeptly handle the complexities of actual degradation patterns.
For an real-world LR test image \( I_{\text{LR}}^{\text{test}} \), the SR network \( \mathcal{S} \) initially produces a super-resolved image \( I_{\text{SR}}^{\text{init}} \). The pre-trained LR reconstruction branch, with its parameters frozen, extracts a degradation embedding \( \mathbf{e}^{\text{test}} \) from \( I_{\text{LR}}^{\text{test}} \), expressed as \( \mathbf{e}^{\text{test}} = \mathcal{E}(I_{\text{LR}}^{\text{test}}) \).
The self-supervised fine-tuning then commences, leveraging \( I_{\text{SR}}^{\text{init}} \) and \( \mathbf{e}^{\text{test}} \) to adjust a specific subset of the SR network's parameters \( \theta_{\text{ft}} \). This fine-tuning is formulated as an optimization problem:
$$
\setlength{\abovedisplayskip}{1pt}
\theta_{\text{ft}}^* = \arg \min_{\theta_{\text{ft}}} \mathcal{L}(\mathcal{R}(\mathcal{S}_{\theta}(I_{\text{LR}}^{\text{test}}), \mathbf{e}^{\text{test}}), I_{\text{LR}}^{\text{test}})\,,
\setlength{\belowdisplayskip}{1pt}
$$
where \( \theta_{\text{ft}}^* \) is the optimized parameters from full model \( \theta \). 

This strategic adjustment enhances the SR network's capability to reconstruct images with high fidelity to the LR inputs, enhances the SR network's ability to generalize to real-world degradation without the need for paired data.

\noindent \textbf{Focused enhancement of high-frequency details.}
%
Conventional SR methods tend to proficiently reconstruct low-frequency regions but often neglect or inadequately restore high-frequency details.
In addition, the low-frequency regions do not require LR reconstruction due to the absence of detailed texture.
Therefore, our approach aims to concentrate the LR reconstruction process specifically on high-frequency areas, thereby preventing the introduction of artifacts into the low-frequency areas.
Specifically, we apply Discrete Wavelet Transform (DWT) to obtain the high-frequency component, and then normalize it to yield a weight map \( \mathbf{W} \in [0, 1] \). This weight map is then utilized to calculate a weighted loss, ensuring the fidelity to high-frequency details:
%
$$
\setlength{\abovedisplayskip}{1pt}
\mathcal{L} = \mathcal{L}_1(\mathbf{W} \odot \hat I_{\text{LR}}^{\text{test}}, \mathbf{W} \odot I_{\text{LR}}^{\text{test}}) + \mathcal{L}_{\text{LPIPS}}(\mathbf{W} \odot \hat I_{\text{LR}}^{\text{test}}, \mathbf{W} \odot I_{\text{LR}}^{\text{test}})\,,
\setlength{\belowdisplayskip}{1pt}
$$
where \( \odot \) denotes element-wise multiplication. The combined loss effectively guides the network to restore high-frequency details with greater precision, improving the perceptual quality of the super-resolved image without compromising low-frequency content.

\begin{figure*}[tbp]
  \centering
  \includegraphics[width=1\linewidth]{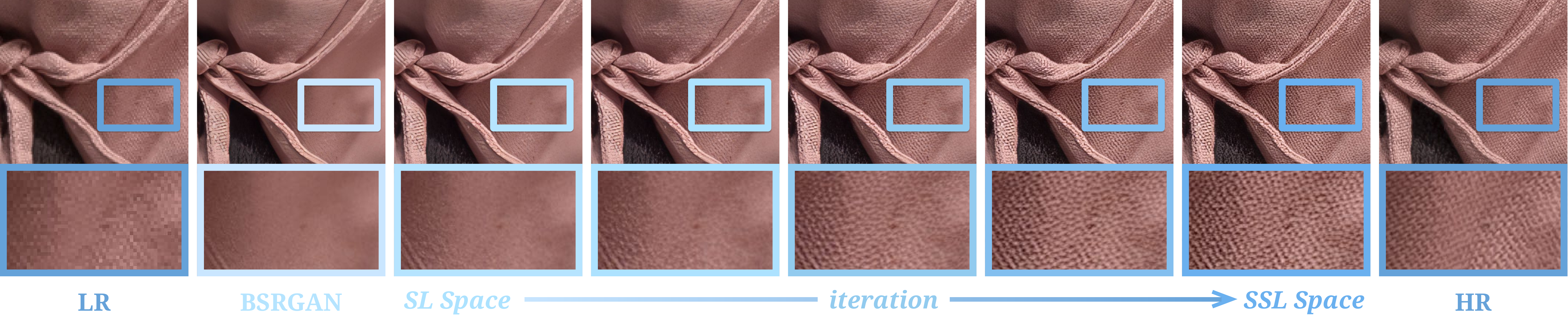} 
  \vspace{-7mm}
  \caption{
  The SR model advances through the proposed fine-tuning iterations, moving from the supervised learning (SL) space of synthetic degradation to the self-supervised learning (SSL) space learned from test images. This results in enhanced SR quality and fidelity.}
  \label{fig:iteration}
  \vspace{-4mm}
\end{figure*}

\vspace{-1mm}
\subsection{Discussion}
\vspace{-1mm}

By combining supervised learning (SL) on synthetic data with self-supervised learning (SSL) on test images with unknown degradation, we dynamically adjust the modeling space based on the intrinsic features of test images, steering the SL space towards a more precise SSL space. 
\figurename~\ref{fig:iteration} shows the effectiveness of our method during the fine-tuning process.
Our method achieves high-quality and high-fidelity SR while maintaining general compatibility across all models. The primary advantages of our approach compared to other methods are included in the following:


\noindent \textbf{General Degradation Modeling.}
The transformation from LR to HR images is recognized as a challenging task, while the reverse HR to LR transformation is comparatively simpler and more robust. Our method capitalizes on this observation, avoiding excessive reliance on extensive paired datasets. Instead, we opt to pre-train a universal degradation embedding extraction and LR reconstruction model. This characteristic ensures that our approach is not bound by assumptions of uniform degradation across image datasets. During the training of the SR model, these parameters remain fixed, allowing the SR model for flexible adaption to unknown distributions in real-world scenarios.
%
%
On the contrary, CycleGAN-based methods simultaneously learn the mappings from LR to HR and HR to LR. This process relies heavily on a substantial amount of data. Furthermore, because CycleGAN implicitly learns the HR to LR mapping without an explicit degradation extraction process, its underlying assumption is that the degradation across the entire dataset is consistent. Consequently, it can only fit certain degradation patterns, largely impacting its performance in real-world scenarios with limited data availability.

\noindent \textbf{Dense pixelwise self-supervision.}
Through self-supervised learning, our method operates independently of external labels, leveraging dense LR pixel-level signals for supervision. This allows the model to learn richer texture features from the intrinsic image structure. This stands in contrast to traditional supervised approaches that rely on discriminators, which may learn inaccurate features due to the sparsity of supervision signals, leading to suboptimal results.

\noindent \textbf{Robust regularization.}
Our approach can be viewed as a form of regularization constraint. 
By integrating degradation embedding extraction and decoupling it from the LR image reconstruction, 
our method maintains effectiveness in guiding the reconstruction process even when faced with imperfect degradation prediction. 
This substantially boosts the robustness of our approach, enabling it to learn rich and accurate texture information from the test images.

\vspace{-2mm}
\section{Experiments}
\vspace{-1mm}

\begin{table*}[]
\centering
\resizebox{\textwidth}{!}{%
\begin{tabular}{@{}c|c|l|ccccccccccccccc@{}}
\toprule \toprule
                           &                             &                           & \multicolumn{6}{c|}{CNN-based}                                                                                                                                                                       & \multicolumn{3}{c|}{Transformer-based}                                                             & \multicolumn{3}{c|}{VQ-based}                                                                  & \multicolumn{3}{c}{Diffusion-based}                                        \\ \cmidrule(l){4-18}
\multirow{-2}{*}{Dataset}  & \multirow{-2}{*}{Sensors}   & \multirow{-2}{*}{Metrics} & {\scriptsize Real-ESRGAN+} & \cellcolor[HTML]{f2f2f2}+ LWay & \multicolumn{1}{c|}{Gain}                               & BSRGAN & \cellcolor[HTML]{f2f2f2}+ LWay & \multicolumn{1}{c|}{Gain}                               & {\footnotesize SwinIR-GAN} & \cellcolor[HTML]{f2f2f2}+ LWay & \multicolumn{1}{c|}{Gain}                               & FeMaSR & \cellcolor[HTML]{f2f2f2}+ LWay & \multicolumn{1}{c|}{Gain}                               & StableSR & \cellcolor[HTML]{f2f2f2}+ LWay & Gain                               \\ \midrule
                           &                             & PSNR {\color[HTML]{369DA2} $\uparrow$}                     & 27.51        & \cellcolor[HTML]{f2f2f2}29.18  & \multicolumn{1}{c|}{\small {\color[HTML]{369DA2} +1.67}}   & 28.81  & \cellcolor[HTML]{f2f2f2}28.85  & \multicolumn{1}{c|}{\small {\color[HTML]{369DA2} +0.04}}   & 28.12      & \cellcolor[HTML]{f2f2f2}28.96  & \multicolumn{1}{c|}{\small {\color[HTML]{369DA2} +0.84}}   & 25.72  & \cellcolor[HTML]{f2f2f2}28.16  & \multicolumn{1}{c|}{\small {\color[HTML]{369DA2} +2.44}}   & 25.50    & \cellcolor[HTML]{f2f2f2}27.22  & \small {\color[HTML]{369DA2} +1.72}   \\
                           &                             & SSIM {\color[HTML]{369DA2} $\uparrow$}                     & 0.8348       & \cellcolor[HTML]{f2f2f2}0.8688 & \multicolumn{1}{c|}{\small {\color[HTML]{369DA2} +0.034}}  & 0.8473 & \cellcolor[HTML]{f2f2f2}0.8496 & \multicolumn{1}{c|}{\small {\color[HTML]{369DA2} +0.0023}} & 0.8486     & \cellcolor[HTML]{f2f2f2}0.8579 & \multicolumn{1}{c|}{\small {\color[HTML]{369DA2} +0.0093}} & 0.7811 & \cellcolor[HTML]{f2f2f2}0.8383 & \multicolumn{1}{c|}{\small {\color[HTML]{369DA2} +0.0572}} & 0.7684   & \cellcolor[HTML]{f2f2f2}0.8043 & \small {\color[HTML]{369DA2} +0.0359} \\
                           &                             & LPIPS {\color[HTML]{369DA2} $\downarrow$}                    & 0.1947       & \cellcolor[HTML]{f2f2f2}0.1479 & \multicolumn{1}{c|}{\small {\color[HTML]{369DA2} -0.0468}} & 0.1988 & \cellcolor[HTML]{f2f2f2}0.1572 & \multicolumn{1}{c|}{\small {\color[HTML]{369DA2} -0.0416}} & 0.1850     & \cellcolor[HTML]{f2f2f2}0.1469 & \multicolumn{1}{c|}{\small {\color[HTML]{369DA2} -0.0381}} & 0.2543 & \cellcolor[HTML]{f2f2f2}0.1747 & \multicolumn{1}{c|}{\small {\color[HTML]{369DA2} -0.0796}} & 0.2636   & \cellcolor[HTML]{f2f2f2}0.2019 & \small {\color[HTML]{369DA2} -0.0617} \\
                           &                             & MAD {\color[HTML]{369DA2} $\downarrow$}                      & 133.96       & \cellcolor[HTML]{f2f2f2}111.91 & \multicolumn{1}{c|}{\small {\color[HTML]{369DA2} -22.05}}  & 119.08 & \cellcolor[HTML]{f2f2f2}116.77 & \multicolumn{1}{c|}{\small {\color[HTML]{369DA2} -2.31}}   & 125.17     & \cellcolor[HTML]{f2f2f2}111.71 & \multicolumn{1}{c|}{\small {\color[HTML]{369DA2} -13.46}}  & 143.38 & \cellcolor[HTML]{f2f2f2}117.48 & \multicolumn{1}{c|}{\small {\color[HTML]{369DA2} -25.90}}  & 145.36   & \cellcolor[HTML]{f2f2f2}124.15 & \small {\color[HTML]{369DA2} -21.21}  \\
                           &                             & NLPD {\color[HTML]{369DA2} $\downarrow$}                     & 0.2807       & \cellcolor[HTML]{f2f2f2}0.2437 & \multicolumn{1}{c|}{\small {\color[HTML]{369DA2} -0.037}}  & 0.2594 & \cellcolor[HTML]{f2f2f2}0.2569 & \multicolumn{1}{c|}{\small {\color[HTML]{369DA2} -0.0025}} & 0.2670     & \cellcolor[HTML]{f2f2f2}0.2541 & \multicolumn{1}{c|}{\small {\color[HTML]{369DA2} -0.0129}} & 0.3239 & \cellcolor[HTML]{f2f2f2}0.2778 & \multicolumn{1}{c|}{\small {\color[HTML]{369DA2} -0.0461}} & 0.3426   & \cellcolor[HTML]{f2f2f2}0.3074 & \small {\color[HTML]{369DA2} -0.0352} \\
                           & \multirow{-6}{*}{Canon}     & DISTIS {\color[HTML]{369DA2} $\downarrow$}                   & 0.1621       & \cellcolor[HTML]{f2f2f2}0.1444 & \multicolumn{1}{c|}{\small {\color[HTML]{369DA2} -0.0177}} & 0.1794 & \cellcolor[HTML]{f2f2f2}0.1558 & \multicolumn{1}{c|}{\small {\color[HTML]{369DA2} -0.0236}} & 0.1557     & \cellcolor[HTML]{f2f2f2}0.1352 & \multicolumn{1}{c|}{\small {\color[HTML]{369DA2} -0.0205}} & 0.2116 & \cellcolor[HTML]{f2f2f2}0.1808 & \multicolumn{1}{c|}{\small {\color[HTML]{369DA2} -0.0308}} & 0.1897   & \cellcolor[HTML]{f2f2f2}0.1596 & \small {\color[HTML]{369DA2} -0.0301} \\ \cmidrule(l){2-18}
                           &                             & PSNR {\color[HTML]{369DA2} $\uparrow$}                     & 26.81        & \cellcolor[HTML]{f2f2f2}28.58  & \multicolumn{1}{c|}{\small {\color[HTML]{369DA2} +1.77}}   & 28.13  & \cellcolor[HTML]{f2f2f2}28.65  & \multicolumn{1}{c|}{\small {\color[HTML]{369DA2} +0.52}}   & 27.54      & \cellcolor[HTML]{f2f2f2}28.55  & \multicolumn{1}{c|}{\small {\color[HTML]{369DA2} +1.01}}   & 25.41  & \cellcolor[HTML]{f2f2f2}27.87  & \multicolumn{1}{c|}{\small {\color[HTML]{369DA2} +2.46}}   & 25.54    & \cellcolor[HTML]{f2f2f2}26.92  & \small {\color[HTML]{369DA2} +1.38}   \\
                           &                             & SSIM {\color[HTML]{369DA2} $\uparrow$}                     & 0.7861       & \cellcolor[HTML]{f2f2f2}0.8249 & \multicolumn{1}{c|}{\small {\color[HTML]{369DA2} +0.0388}} & 0.8012 & \cellcolor[HTML]{f2f2f2}0.8057 & \multicolumn{1}{c|}{\small {\color[HTML]{369DA2} +0.0045}} & 0.8043     & \cellcolor[HTML]{f2f2f2}0.813  & \multicolumn{1}{c|}{\small {\color[HTML]{369DA2} +0.0087}} & 0.7314 & \cellcolor[HTML]{f2f2f2}0.7936 & \multicolumn{1}{c|}{\small {\color[HTML]{369DA2} +0.0622}} & 0.7370   & \cellcolor[HTML]{f2f2f2}0.7686 & \small {\color[HTML]{369DA2} +0.0316} \\
                           &                             & LPIPS {\color[HTML]{369DA2} $\downarrow$}                    & 0.2300       & \cellcolor[HTML]{f2f2f2}0.1769 & \multicolumn{1}{c|}{\small {\color[HTML]{369DA2} -0.0531}} & 0.2302 & \cellcolor[HTML]{f2f2f2}0.1750 & \multicolumn{1}{c|}{\small {\color[HTML]{369DA2} -0.0552}} & 0.2154     & \cellcolor[HTML]{f2f2f2}0.176  & \multicolumn{1}{c|}{\small {\color[HTML]{369DA2} -0.0394}} & 0.2738 & \cellcolor[HTML]{f2f2f2}0.2028 & \multicolumn{1}{c|}{\small {\color[HTML]{369DA2} -0.071}}  & 0.2711   & \cellcolor[HTML]{f2f2f2}0.2156 & \small {\color[HTML]{369DA2} -0.0555} \\
                           &                             & MAD {\color[HTML]{369DA2} $\downarrow$}                      & 131.62       & \cellcolor[HTML]{f2f2f2}108.18 & \multicolumn{1}{c|}{\small {\color[HTML]{369DA2} -23.44}}  & 118.48 & \cellcolor[HTML]{f2f2f2}105.64 & \multicolumn{1}{c|}{\small {\color[HTML]{369DA2} -12.84}}  & 122.65     & \cellcolor[HTML]{f2f2f2}106.73 & \multicolumn{1}{c|}{\small {\color[HTML]{369DA2} -15.92}}  & 137.54 & \cellcolor[HTML]{f2f2f2}110.79 & \multicolumn{1}{c|}{\small {\color[HTML]{369DA2} -26.75}}  & 139.26   & \cellcolor[HTML]{f2f2f2}119.29 & \small {\color[HTML]{369DA2} -19.97}  \\
                           &                             & NLPD {\color[HTML]{369DA2} $\downarrow$}                     & 0.3061       & \cellcolor[HTML]{f2f2f2}0.2667 & \multicolumn{1}{c|}{\small {\color[HTML]{369DA2} -0.0394}} & 0.2805 & \cellcolor[HTML]{f2f2f2}0.2758 & \multicolumn{1}{c|}{\small {\color[HTML]{369DA2} -0.0047}} & 0.2844     & \cellcolor[HTML]{f2f2f2}0.272  & \multicolumn{1}{c|}{\small {\color[HTML]{369DA2} -0.0124}} & 0.3419 & \cellcolor[HTML]{f2f2f2}0.297  & \multicolumn{1}{c|}{\small {\color[HTML]{369DA2} -0.0449}} & 0.3513   & \cellcolor[HTML]{f2f2f2}0.3215 & \small {\color[HTML]{369DA2} -0.0298} \\
\multirow{-12}{*}{RealSR}  & \multirow{-6}{*}{Nikon}     & DISTIS {\color[HTML]{369DA2} $\downarrow$}                   & 0.1950       & \cellcolor[HTML]{f2f2f2}0.1714 & \multicolumn{1}{c|}{\small {\color[HTML]{369DA2} -0.0236}} & 0.2102 & \cellcolor[HTML]{f2f2f2}0.1791 & \multicolumn{1}{c|}{\small {\color[HTML]{369DA2} -0.0311}} & 0.1842     & \cellcolor[HTML]{f2f2f2}0.1639 & \multicolumn{1}{c|}{\small {\color[HTML]{369DA2} -0.0203}} & 0.2340 & \cellcolor[HTML]{f2f2f2}0.2042 & \multicolumn{1}{c|}{\small {\color[HTML]{369DA2} -0.0298}} & 0.2131   & \cellcolor[HTML]{f2f2f2}0.1837 & \small {\color[HTML]{369DA2} -0.0294} \\ \midrule \midrule
                           &                             & PSNR {\color[HTML]{369DA2} $\uparrow$}                     & 30.16        & \cellcolor[HTML]{f2f2f2}31.4   & \multicolumn{1}{c|}{\small {\color[HTML]{369DA2} +1.24}}   & 30.47  & \cellcolor[HTML]{f2f2f2}31.23  & \multicolumn{1}{c|}{\small {\color[HTML]{369DA2} +0.76}}   & 29.92      & \cellcolor[HTML]{f2f2f2}30.77  & \multicolumn{1}{c|}{\small {\color[HTML]{369DA2} +0.85}}   & 27.51  & \cellcolor[HTML]{f2f2f2}29.75  & \multicolumn{1}{c|}{\small {\color[HTML]{369DA2} +2.24}}   & 28.63    & \cellcolor[HTML]{f2f2f2}29.28  & \small {\color[HTML]{369DA2} +0.65}   \\
                           &                             & SSIM {\color[HTML]{369DA2} $\uparrow$}                     & 0.8326       & \cellcolor[HTML]{f2f2f2}0.8597 & \multicolumn{1}{c|}{\small {\color[HTML]{369DA2} +0.0271}} & 0.8260 & \cellcolor[HTML]{f2f2f2}0.8442 & \multicolumn{1}{c|}{\small {\color[HTML]{369DA2} +0.0182}} & 0.8213     & \cellcolor[HTML]{f2f2f2}0.8398 & \multicolumn{1}{c|}{\small {\color[HTML]{369DA2} +0.0185}} & 0.7725 & \cellcolor[HTML]{f2f2f2}0.8096 & \multicolumn{1}{c|}{\small {\color[HTML]{369DA2} +0.0371}} & 0.7648   & \cellcolor[HTML]{f2f2f2}0.7785 & \small {\color[HTML]{369DA2} +0.0137} \\
                           &                             & LPIPS {\color[HTML]{369DA2} $\downarrow$}                    & 0.2488       & \cellcolor[HTML]{f2f2f2}0.2341 & \multicolumn{1}{c|}{\small {\color[HTML]{369DA2} -0.0147}} & 0.2685 & \cellcolor[HTML]{f2f2f2}0.2469 & \multicolumn{1}{c|}{\small {\color[HTML]{369DA2} -0.0216}} & 0.2565     & \cellcolor[HTML]{f2f2f2}0.2383 & \multicolumn{1}{c|}{\small {\color[HTML]{369DA2} -0.0182}} & 0.3228 & \cellcolor[HTML]{f2f2f2}0.2931 & \multicolumn{1}{c|}{\small {\color[HTML]{369DA2} -0.0297}} & 0.3331   & \cellcolor[HTML]{f2f2f2}0.3017 & \small {\color[HTML]{369DA2} -0.0314} \\
                           &                             & MAD {\color[HTML]{369DA2} $\downarrow$}                      & 125.20       & \cellcolor[HTML]{f2f2f2}112.1  & \multicolumn{1}{c|}{\small {\color[HTML]{369DA2} -13.10}}  & 123.22 & \cellcolor[HTML]{f2f2f2}115.14 & \multicolumn{1}{c|}{\small {\color[HTML]{369DA2} -8.08}}   & 124.85     & \cellcolor[HTML]{f2f2f2}114.09 & \multicolumn{1}{c|}{\small {\color[HTML]{369DA2} -10.76}}  & 140.50 & \cellcolor[HTML]{f2f2f2}125.52 & \multicolumn{1}{c|}{\small {\color[HTML]{369DA2} -14.98}}  & 141.13   & \cellcolor[HTML]{f2f2f2}130.01 & \small {\color[HTML]{369DA2} -11.12}  \\
                           &                             & NLPD {\color[HTML]{369DA2} $\downarrow$}                     & 0.3032       & \cellcolor[HTML]{f2f2f2}0.2751 & \multicolumn{1}{c|}{\small {\color[HTML]{369DA2} -0.0281}} & 0.3034 & \cellcolor[HTML]{f2f2f2}0.2857 & \multicolumn{1}{c|}{\small {\color[HTML]{369DA2} -0.0177}} & 0.3105     & \cellcolor[HTML]{f2f2f2}0.2895 & \multicolumn{1}{c|}{\small {\color[HTML]{369DA2} -0.021}}  & 0.3502 & \cellcolor[HTML]{f2f2f2}0.3152 & \multicolumn{1}{c|}{\small {\color[HTML]{369DA2} -0.035}}  & 0.3503   & \cellcolor[HTML]{f2f2f2}0.3402 & \small {\color[HTML]{369DA2} -0.0101} \\
                           & \multirow{-6}{*}{sony}      & DISTIS {\color[HTML]{369DA2} $\downarrow$}                   & 0.1859       & \cellcolor[HTML]{f2f2f2}0.1765 & \multicolumn{1}{c|}{\small {\color[HTML]{369DA2} -0.0094}} & 0.2115 & \cellcolor[HTML]{f2f2f2}0.1934 & \multicolumn{1}{c|}{\small {\color[HTML]{369DA2} -0.0181}} & 0.1883     & \cellcolor[HTML]{f2f2f2}0.1783 & \multicolumn{1}{c|}{\small {\color[HTML]{369DA2} -0.01}}   & 0.2314 & \cellcolor[HTML]{f2f2f2}0.2168 & \multicolumn{1}{c|}{\small {\color[HTML]{369DA2} -0.0146}} & 0.2296   & \cellcolor[HTML]{f2f2f2}0.2176 & \small {\color[HTML]{369DA2} -0.012}  \\ \cmidrule(l){2-18}
                           &                             & PSNR {\color[HTML]{369DA2} $\uparrow$}                     & 29.53        & \cellcolor[HTML]{f2f2f2}29.88  & \multicolumn{1}{c|}{\small {\color[HTML]{369DA2} +0.35}}   & 29.16  & \cellcolor[HTML]{f2f2f2}29.4   & \multicolumn{1}{c|}{\small {\color[HTML]{369DA2} +0.24}}   & 28.94      & \cellcolor[HTML]{f2f2f2}29.57  & \multicolumn{1}{c|}{\small {\color[HTML]{369DA2} +0.63}}   & 26.42  & \cellcolor[HTML]{f2f2f2}28.26  & \multicolumn{1}{c|}{\small {\color[HTML]{369DA2} +1.84}}   & 28.69    & \cellcolor[HTML]{f2f2f2}29.05  & \small {\color[HTML]{369DA2} +0.36}   \\
                           &                             & SSIM {\color[HTML]{369DA2} $\uparrow$}                     & 0.8050       & \cellcolor[HTML]{f2f2f2}0.8206 & \multicolumn{1}{c|}{\small {\color[HTML]{369DA2} +0.0156}} & 0.7931 & \cellcolor[HTML]{f2f2f2}0.7944 & \multicolumn{1}{c|}{\small {\color[HTML]{369DA2} +0.0013}} & 0.8002     & \cellcolor[HTML]{f2f2f2}0.8071 & \multicolumn{1}{c|}{\small {\color[HTML]{369DA2} +0.0069}} & 0.6976 & \cellcolor[HTML]{f2f2f2}0.7557 & \multicolumn{1}{c|}{\small {\color[HTML]{369DA2} +0.0581}} & 0.7460   & \cellcolor[HTML]{f2f2f2}0.7487 & \small {\color[HTML]{369DA2} +0.0027} \\
                           &                             & LPIPS {\color[HTML]{369DA2} $\downarrow$}                    & 0.3107       & \cellcolor[HTML]{f2f2f2}0.308  & \multicolumn{1}{c|}{\small {\color[HTML]{369DA2} -0.0027}} & 0.3275 & \cellcolor[HTML]{f2f2f2}0.2926 & \multicolumn{1}{c|}{\small {\color[HTML]{369DA2} -0.0349}} & 0.3184     & \cellcolor[HTML]{f2f2f2}0.3093 & \multicolumn{1}{c|}{\small {\color[HTML]{369DA2} -0.0091}} & 0.4129 & \cellcolor[HTML]{f2f2f2}0.3762 & \multicolumn{1}{c|}{\small {\color[HTML]{369DA2} -0.0367}} & 0.3853   & \cellcolor[HTML]{f2f2f2}0.3800 & \small {\color[HTML]{369DA2} -0.0053} \\
                           &                             & MAD {\color[HTML]{369DA2} $\downarrow$}                      & 127.91       & \cellcolor[HTML]{f2f2f2}125.04 & \multicolumn{1}{c|}{\small {\color[HTML]{369DA2} -2.87}}   & 130.94 & \cellcolor[HTML]{f2f2f2}126.87 & \multicolumn{1}{c|}{\small {\color[HTML]{369DA2} -4.07}}   & 131.73     & \cellcolor[HTML]{f2f2f2}126.04 & \multicolumn{1}{c|}{\small {\color[HTML]{369DA2} -5.69}}   & 151.35 & \cellcolor[HTML]{f2f2f2}138.85 & \multicolumn{1}{c|}{\small {\color[HTML]{369DA2} -12.50}}  & 137.60   & \cellcolor[HTML]{f2f2f2}132.71 & \small {\color[HTML]{369DA2} -4.89}   \\
                           &                             & NLPD {\color[HTML]{369DA2} $\downarrow$}                     & 0.3016       & \cellcolor[HTML]{f2f2f2}0.2899 & \multicolumn{1}{c|}{\small {\color[HTML]{369DA2} -0.0117}} & 0.3157 & \cellcolor[HTML]{f2f2f2}0.3129 & \multicolumn{1}{c|}{\small {\color[HTML]{369DA2} -0.0028}} & 0.3093     & \cellcolor[HTML]{f2f2f2}0.3005 & \multicolumn{1}{c|}{\small {\color[HTML]{369DA2} -0.0088}} & 0.3897 & \cellcolor[HTML]{f2f2f2}0.3425 & \multicolumn{1}{c|}{\small {\color[HTML]{369DA2} -0.0472}} & 0.3410   & \cellcolor[HTML]{f2f2f2}0.3353 & \small {\color[HTML]{369DA2} -0.0057} \\
                           & \multirow{-6}{*}{olympus}   & DISTIS {\color[HTML]{369DA2} $\downarrow$}                   & 0.2130       & \cellcolor[HTML]{f2f2f2}0.2118 & \multicolumn{1}{c|}{\small {\color[HTML]{369DA2} -0.0012}} & 0.2276 & \cellcolor[HTML]{f2f2f2}0.2145 & \multicolumn{1}{c|}{\small {\color[HTML]{369DA2} -0.0131}} & 0.2181     & \cellcolor[HTML]{f2f2f2}0.2109 & \multicolumn{1}{c|}{\small {\color[HTML]{369DA2} -0.0072}} & 0.2552 & \cellcolor[HTML]{f2f2f2}0.2406 & \multicolumn{1}{c|}{\small {\color[HTML]{369DA2} -0.0146}} & 0.2412   & \cellcolor[HTML]{f2f2f2}0.2371 & \small {\color[HTML]{369DA2} -0.0041} \\ \cmidrule(l){2-18}                        
                           &                             & PSNR {\color[HTML]{369DA2} $\uparrow$}                     & 29.81        & \cellcolor[HTML]{f2f2f2}30.83  & \multicolumn{1}{c|}{\small {\color[HTML]{369DA2} +1.02}}   & 29.98  & \cellcolor[HTML]{f2f2f2}31.05  & \multicolumn{1}{c|}{\small {\color[HTML]{369DA2} +1.07}}   & 29.11      & \cellcolor[HTML]{f2f2f2}30.94  & \multicolumn{1}{c|}{\small {\color[HTML]{369DA2} +1.83}}   & 27.83  & \cellcolor[HTML]{f2f2f2}29.44  & \multicolumn{1}{c|}{\small {\color[HTML]{369DA2} +1.61}}   & 29.13    & \cellcolor[HTML]{f2f2f2}29.88  & \small {\color[HTML]{369DA2} +0.75}   \\
                           &                             & SSIM {\color[HTML]{369DA2} $\uparrow$}                     & 0.8094       & \cellcolor[HTML]{f2f2f2}0.8283 & \multicolumn{1}{c|}{\small {\color[HTML]{369DA2} +0.0189}} & 0.7987 & \cellcolor[HTML]{f2f2f2}0.8236 & \multicolumn{1}{c|}{\small {\color[HTML]{369DA2} +0.0249}} & 0.7918     & \cellcolor[HTML]{f2f2f2}0.8193 & \multicolumn{1}{c|}{\small {\color[HTML]{369DA2} +0.0275}} & 0.7413 & \cellcolor[HTML]{f2f2f2}0.7798 & \multicolumn{1}{c|}{\small {\color[HTML]{369DA2} +0.0385}} & 0.7428   & \cellcolor[HTML]{f2f2f2}0.7554 & \small {\color[HTML]{369DA2} +0.0126} \\
                           &                             & LPIPS {\color[HTML]{369DA2} $\downarrow$}                    & 0.2592       & \cellcolor[HTML]{f2f2f2}0.2581 & \multicolumn{1}{c|}{\small {\color[HTML]{369DA2} -0.0011}} & 0.2738 & \cellcolor[HTML]{f2f2f2}0.2624 & \multicolumn{1}{c|}{\small {\color[HTML]{369DA2} -0.0114}} & 0.2688     & \cellcolor[HTML]{f2f2f2}0.2517 & \multicolumn{1}{c|}{\small {\color[HTML]{369DA2} -0.0171}} & 0.3144 & \cellcolor[HTML]{f2f2f2}0.2973 & \multicolumn{1}{c|}{\small {\color[HTML]{369DA2} -0.0171}} & 0.3143   & \cellcolor[HTML]{f2f2f2}0.3021 & \small {\color[HTML]{369DA2} -0.0122} \\
                           &                             & MAD {\color[HTML]{369DA2} $\downarrow$}                      & 124.51       & \cellcolor[HTML]{f2f2f2}116.18 & \multicolumn{1}{c|}{\small {\color[HTML]{369DA2} -8.33}}   & 124.38 & \cellcolor[HTML]{f2f2f2}114.04 & \multicolumn{1}{c|}{\small {\color[HTML]{369DA2} -10.34}}  & 126.61     & \cellcolor[HTML]{f2f2f2}112.79 & \multicolumn{1}{c|}{\small {\color[HTML]{369DA2} -13.82}}  & 137.50 & \cellcolor[HTML]{f2f2f2}124.81 & \multicolumn{1}{c|}{\small {\color[HTML]{369DA2} -12.69}}  & 132.36   & \cellcolor[HTML]{f2f2f2}122.85 & \small {\color[HTML]{369DA2} -9.51}   \\
                           &                             & NLPD {\color[HTML]{369DA2} $\downarrow$}                     & 0.304        & \cellcolor[HTML]{f2f2f2}0.2825 & \multicolumn{1}{c|}{\small {\color[HTML]{369DA2} -0.0215}} & 0.3109 & \cellcolor[HTML]{f2f2f2}0.2852 & \multicolumn{1}{c|}{\small {\color[HTML]{369DA2} -0.0257}} & 0.3184     & \cellcolor[HTML]{f2f2f2}0.2869 & \multicolumn{1}{c|}{\small {\color[HTML]{369DA2} -0.0315}} & 0.3604 & \cellcolor[HTML]{f2f2f2}0.3215 & \multicolumn{1}{c|}{\small {\color[HTML]{369DA2} -0.0389}} & 0.3444   & \cellcolor[HTML]{f2f2f2}0.3312 & \small {\color[HTML]{369DA2} -0.0132} \\
\multirow{-18}{*}{DRealSR} & \multirow{-6}{*}{panasonic} & DISTIS {\color[HTML]{369DA2} $\downarrow$}                   & 0.2000       & \cellcolor[HTML]{f2f2f2}0.1974 & \multicolumn{1}{c|}{\small {\color[HTML]{369DA2} -0.0026}} & 0.2130 & \cellcolor[HTML]{f2f2f2}0.2021 & \multicolumn{1}{c|}{\small {\color[HTML]{369DA2} -0.0109}} & 0.2046     & \cellcolor[HTML]{f2f2f2}0.1948 & \multicolumn{1}{c|}{\small {\color[HTML]{369DA2} -0.0098}} & 0.2243 & \cellcolor[HTML]{f2f2f2}0.2121 & \multicolumn{1}{c|}{\small {\color[HTML]{369DA2} -0.0122}} & 0.2255   & \cellcolor[HTML]{f2f2f2}0.2196 & \small {\color[HTML]{369DA2} -0.0059} \\ 

\bottomrule \bottomrule
\end{tabular}%
}
\vspace{-3mm}
\caption{The performance improvements across various model types utilizing our proposed training methodology.}
\vspace{-3mm}
\label{tab:main}
\end{table*}

\begin{table}[t]
\centering
\resizebox{\linewidth}{!}{%
\begin{tabular}{cccccc}
\toprule \toprule
\begin{tabular}[c]{@{}c@{}}
\# of Fine-tuning \\ Images Per Model\end{tabular} &
  Description &
  LPIPS {\color[HTML]{369DA2} $\downarrow$} &
  DISTIS {\color[HTML]{369DA2} $\downarrow$} &
  MAD {\color[HTML]{369DA2} $\downarrow$}
  \\ \midrule  \midrule 
0  & \begin{tabular}[c]{@{}c@{}}baseline,  \\ without fine-tuning\end{tabular}        & 0.3136 & 0.2353 & 117.71   \\ \midrule 
1  & \begin{tabular}[c]{@{}c@{}} fine-tuning \\ on every single images\end{tabular}   & \textbf{0.2351} & \textbf{0.1919} & 111.46   \\  \midrule 
10 & \begin{tabular}[c]{@{}c@{}} fine-tuning \\ on the entire testset\end{tabular}    & 0.2536 & 0.2044 & 111.63   \\ \midrule 
50 &
  \begin{tabular}[c]{@{}c@{}} fine-tuning with 40 additional\\ images from the same sensors\end{tabular} & 
  0.2571 &
  0.2037 &
  \textbf{108.62} \\
  \bottomrule \bottomrule  
\end{tabular}%
}
\vspace{-2mm}
\caption{The impact of the number of images used for a single fine-tuning training. Our method can be fine-tuned either on individual images or on the entire test set, which greatly reduces cost.
}\label{tab:ablation2}
\vspace{-5mm}
\end{table}

\subsection{Experimental Settings}
\vspace{-1mm}

\noindent \textbf{Testing methods.}
%
%
%
Our proposed method serves as a universally applicable self-supervised learning strategy for various cutting-edge blind SR models, eliminating the necessity for architectural modifications. We conduct evaluations on a diverse range of advanced SR methods, including BSRGAN~\cite{zhang2021designing} and Real-ESRGAN+~\cite{wang2021real} employing conventional CNN frameworks, SwinIR-GAN~\cite{liang2021swinir} integrating Transformer structures, FeMaSR~\cite{chen2022real} utilizing VQGAN, and StableSR~\cite{stablesr} based on pre-trained diffusion. We use officially released SR models as baselines and conduct self-supervised fine-tuning on targeted test datasets. While fine-tuning a single image can lead to superior performance, for improved training efficiency, we opt to fine-tune the entire test dataset collectively. All experiments are conducted under this configuration unless otherwise specified.

%

\noindent \textbf{Implementation details.}
%
%
%
%
We adopt the Adam~\cite{KingBa15} optimizer. For StableSR, we set the learning rate to 5e-5 and the batch size to 1. For the remaining models, a learning rate of 2e-6 and a batch size of 6 are used. Each model undergoes rapid fine-tuning on a single V100 GPU. The duration of training varies among models and images, typically spanning 150 to 500 iterations. \textit{More details are provided in the supplementary materials.}

\noindent \textbf{Training datasets.} 
Our self-supervised fine-tuning approach is directly applied to the test set, without the need for a separate training set. The only prerequisite training is allocated for the LR reconstruction network, which is trained using 6,000 real paired images collected in-house. It is critical to note that these data were invisible to the SR network.

\noindent \textbf{Testing datasets.}
%
Our method is evaluated on real-world paired datasets, including RealSR~\cite{cai2019toward} and DRealSR~\cite{wei2020component}. These datasets are meticulously curated from diverse device sensors to reflect various degradation characteristics. To ensure a fair comparison with other methods, we follow the standard setting of cropping each image into multiple patches for a 4$\times$ SR. The LR image patch size is 128 $\times$ 128, while the corresponding HR size is 512 $\times$ 512.

\noindent \textbf{Evaluation metrics.}
%
%
%
%
We employ LPIPS~\cite{zhang2018unreasonable}, DISTIS~\cite{ding2020image}, and NLPD~\cite{hepburn2019enforcing} metrics that closely align with human perception~\cite{jinjin2020pipal, gu2020image}. Additionally, traditional metrics such as PSNR, SSIM \cite{wang2004image}, and MAD \cite{larson2010most} are included for a comprehensive assessment.
Six different metrics provide a comprehensive evaluation.

\vspace{-2mm}
\subsection{Improvements on Existing Methods}
\vspace{-2mm}

%
%
%
%
The results outlined in \tablename~\ref{tab:main} compellingly demonstrate our method's effectiveness in significantly advancing SR quality. Notably, improvements are consistently observed across all models, datasets, and metrics, underscoring the universal applicability of our approach. 
For CNN-based models like Real-ESRGAN+, our method achieves a notable enhancement on the Nikon dataset, delivering a 1.77dB improvement in PSNR and a 0.0388 increase in SSIM. These improvements contribute to more precise reconstruction of high-quality images. Furthermore, the validation of enhanced perceptual quality is evident through an LPIPS reduction of 0.0532. Additionally, when applied to Transformer models such as SwinIR-GAN, our method showcases considerable improvements. On the Olympus dataset, we observe a 0.63 dB increase in PSNR and a significant decrease in MAD by 5.69, highlighting the framework's capacity to enhance fidelity and sharpness.

\begin{figure*}[tbp]
  \centering
  \includegraphics[width=1\linewidth]{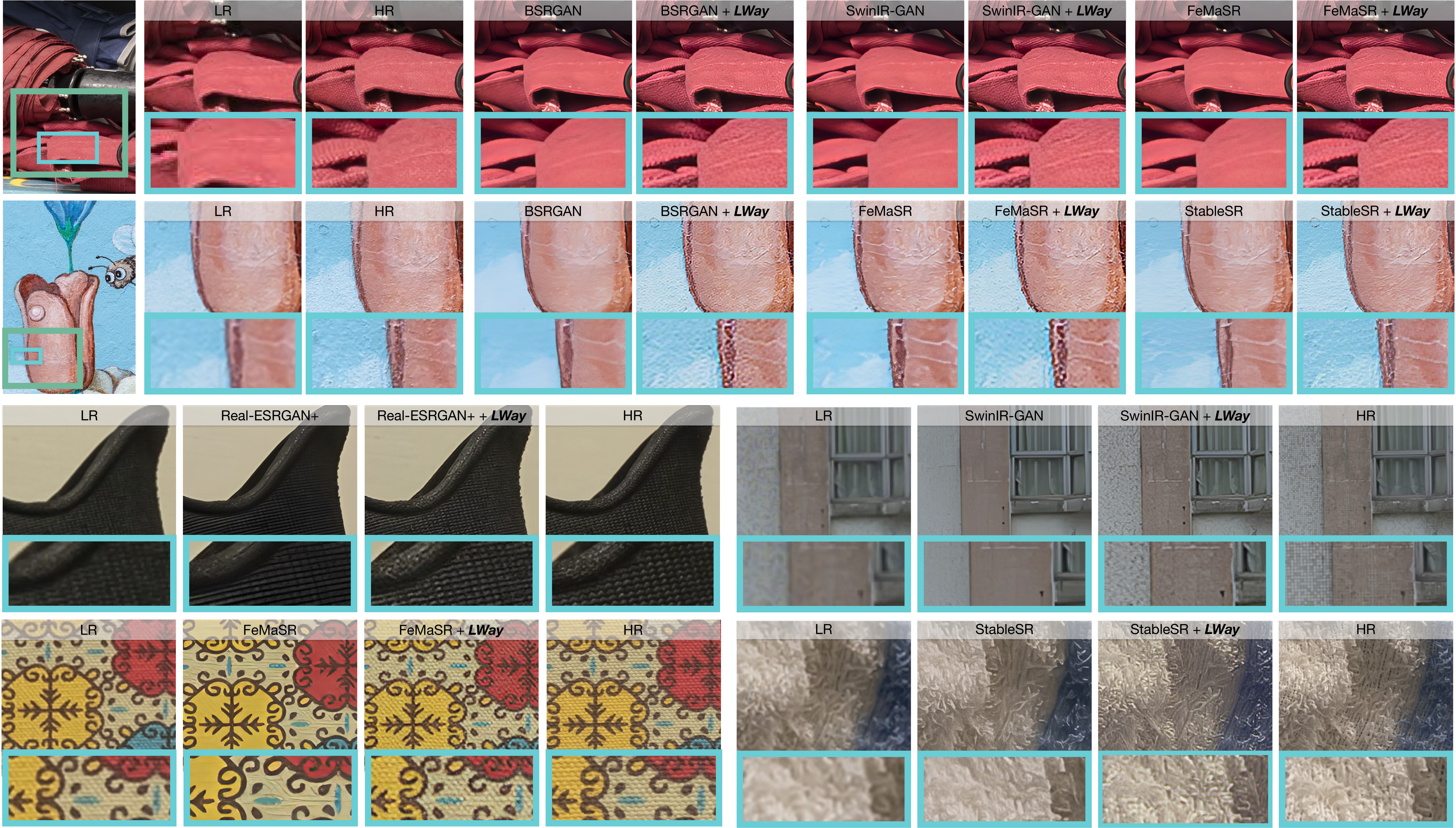} 
  \vspace{-7mm}
  \caption{Qualitative comparisons on real-world datasets. The content within the {\color[HTML]{369DA2} blue box} represents a zoomed-in image.}
  \vspace{-3mm}  
  \label{fig:visual}
\end{figure*}

%
As depicted in \figurename~\ref{fig:visual}, in the first example, all SR models fail to preserve the original textures present in the input images, resulting in excessively smoothed fabric patterns. However, upon applying our self-supervised fine-tuning method, significant improvements are observed across all approaches, successfully reconstructing clear fabric textures. A similar improvement is evident in the second example of oil paintings. The existing SR models struggle to capture the intricate details of the paintings. Conversely, our method effectively restores the artistic effects, particularly showcasing notable enhancement for StableSR.
For other examples, the results are similar as well, our method significantly improving high-frequency detail recovery, yielding results that were both sharp and rich in detail.

\begin{figure*}[tbp]
  \centering
  \includegraphics[width=1\linewidth]{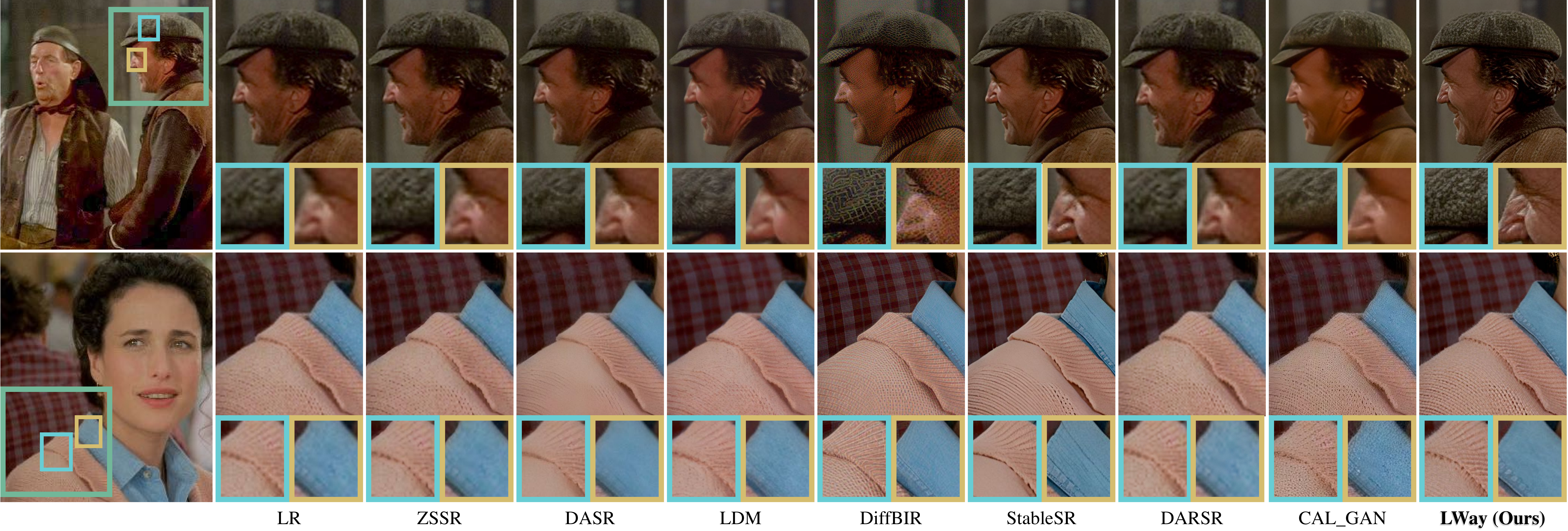} 
  \vspace{-7mm}  
  \caption{Qualitative comparisons on two old films.}
  \vspace{-3mm}  
  \label{fig:film}
\end{figure*}

\begin{figure*}[!tbp]
  \centering
  \includegraphics[width=1\linewidth]{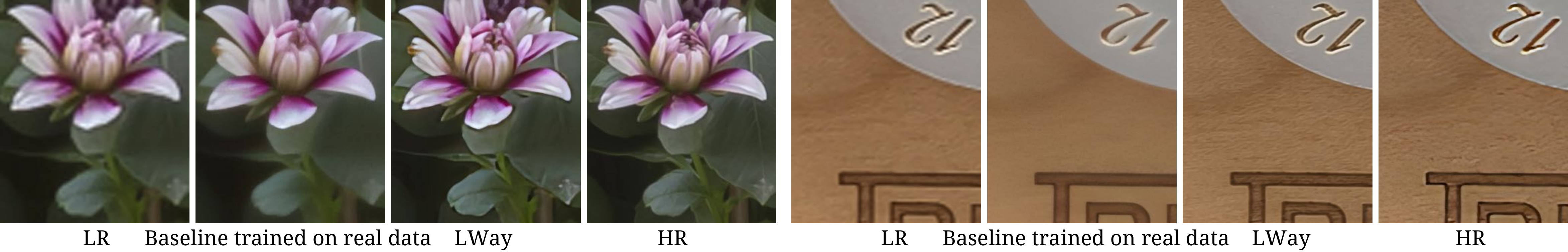} 
  \vspace{-6mm}
  \caption{Supervised fine-tuning a baseline model on one real dataset doesn't perform well on another due to dataset gaps. 
  Our proposed method self-supervised fine-tuned model for specific test images achieves superior performance.
  }
  \label{fig:abaltionvisual}
\vspace{-4mm}  
\end{figure*}

\vspace{-2mm}
\subsection{Application on Real-world Scenes}
\vspace{-2mm}

%
Old films often exhibit issues like graininess, color fading, and lower resolution, making them an ideal testbed for evaluating the practical capabilities of SR models. To conduct a comprehensive comparison, we curate a selection of state-of-the-art real-world SR models. These encompass various methodologies: ZSSR~\cite{shocher2018zero}, a self-supervised learning model; DASR~\cite{liang2022efficient}, a degradation-adaptive approach; large diffusion models such as LDM~\cite{rombach2022high}, DiffBIR~\cite{lin2023diffbir}, and StableSR~\cite{stablesr}; DARSR~\cite{zhou2023learning}, which leverages unsupervised techniques for enhanced model performance; and CAL\_GAN~\cite{park2023content}, a photo-realistic SR model. We employ StableSR as the base model and implement the proposed self-supervised learning strategy. The first case in Figure~\ref{fig:film} involves a 480p low-resolution film, namely ``My Fair Lady''. Among the assessed models, ZSSR, DASR, and DARSR exhibit minimal improvements, while DiffBIR introduces unpleasing artifacts. Other models achieve slightly smoother results. Notably, our model not only accurately reproduces the hat with clear fabric textures but also effectively restores facial features, including wrinkles and contours. In contrast to some methods that may introduce unnatural effects or overly smooth distortions, our model adeptly balances the restoration of fine textures with preserving overall image clarity.

\noindent \textbf{User study.} We conducted a user study with the participation of 24 experienced researchers. Each participant was tasked with assigning a visual perceptual quality score ranging from 0 to 10 to every image. The results, depicted in the \figurename~\ref{fig:userstudy}, reveal a significant lead of our proposed method over alternative approaches, surpassing the second-best method by more than 2 points. Notably, the scores for DASR, DiffBIR, and DARSR were even lower than those for LR images, indicating a limited effectiveness of these methods in handling real-world images.

\begin{table}[t]
\centering
\resizebox{1\linewidth}{!}{%
\begin{tabular}{lcccc}
\toprule 

Training Type &
  \begin{tabular}[c]{@{}c@{}}\quad Number of \quad\\ \quad Sensors\quad\end{tabular} &
  \begin{tabular}[c]{@{}c@{}}\quad Number of \quad\\ \quad Images\quad\end{tabular} & 
  \quad LPIPS  {\color[HTML]{369DA2} $\downarrow$} \quad&
  \quad DISTIS {\color[HTML]{369DA2} $\downarrow$} \quad
  \\ \midrule 
- (baseline)                     & - & -    & 0.2302 & 0.2102      \\ \midrule
\multirow{2}{*}{Synthetic Data}  & - & 2K  & 0.1836 & 0.1885      \\ 
                                 & - & 6k   & 0.1816 & 0.1873      \\ \midrule
\multirow{4}{*}{Real-world Data} & 1 & 0.6K & 0.2003 & 0.1970      \\
                                 & 2 & 2K   & 0.1785 & 0.1793      \\
                                 & 2 & 4K   & \textbf{0.1722} & \textbf{0.1772}      \\
                                 & 3 & 6K   & 0.1800 & 0.1830    \\
\bottomrule
\end{tabular}%
}
\vspace{-2mm}
\caption{Ablation on training data of LR reconstruction.}\label{tab:ablation1}
\vspace{-1mm}
\end{table}

\begin{table}[t]
\centering
\resizebox{1\linewidth}{!}{%
\begin{tabular}{@{}l *{7}{>{\centering\arraybackslash}p{1.4cm}}@{}}
\toprule
   & baseline & 128    & 256    & 512    & 1024   & \textbf{2048}   & 4096   \\ \midrule 
PSNR{\color[HTML]{369DA2} $\uparrow$}    & 28.13    & 28.92  & 28.54  & 28.85  & 29.10  & \textbf{29.56}  & 29.20  \\
LPIPS{\color[HTML]{369DA2} $\downarrow$} & 0.2302   & 0.1804 & 0.1776 & 0.1722 & 0.1736 & \textbf{0.1629} & 0.1669 \\
DISTS{\color[HTML]{369DA2} $\downarrow$} & 0.2192   & 0.1792 & 0.1818 & 0.1772 & 0.1749 & \textbf{0.1630} & 0.1656 \\ \bottomrule
\end{tabular}
}
\vspace{-2mm}
\caption{Ablation on dimensions of degradation embedding.}\label{tab:dimension}
\vspace{-1mm}
\end{table}

\begin{table}[tbp]
  \centering
  \begin{minipage}[t]{0.538\linewidth}
    \centering
    \resizebox{\textwidth}{!}{%
      \begin{tabular}{lcc}
        \toprule
        Method & LPIPS {\color[HTML]{369DA2} $\downarrow$} & DISTIS {\color[HTML]{369DA2} $\downarrow$} \\
        \midrule 
        baseline               & 0.2302 & 0.2102      \\
        baseline + real data   & 0.2268 & 0.1989      \\
        \textbf{LWay (ours)}   & \textbf{0.1722} & \textbf{0.1772}   \\
        \bottomrule
      \end{tabular}%
    }
    \vspace{-2mm}
    \caption{Our method versus supervised real data fine-tuning.}\label{tab:ablationdata}
  \end{minipage}
  \hfill
  \begin{minipage}[t]{0.42\linewidth}
    \centering
    \resizebox{\textwidth}{!}{%
      \begin{tabular}{ccc}
        \toprule
        HF Loss  & LPIPS {\color[HTML]{369DA2} $\downarrow$} & DISTIS {\color[HTML]{369DA2} $\downarrow$}\\
        \midrule  
        \scalebox{0.75}{\usym{2613}}  & 0.1858 & 0.1879  \\ 
        \scalebox{0.75}{\usym{1F5F8}} & 0.1722 & 0.1772  \\  \bottomrule
                                      &        &      
      \end{tabular}%
    }
    \vspace{-2mm}    
    \caption{Ablation study on high-frequency (HF) loss.}\label{tab:ablationloss}
  \end{minipage}
\vspace{-3mm}
\end{table}

\subsection{Ablation Study}
\vspace{-1mm}
We conducted an ablation study on the RealSR Nikon test set using BSRGAN. We trained 65\% of the model parameters to achieve the lowest LPIPS score on this test set.

\begin{figure}[tbp]
  \centering
  \begin{minipage}[t]{0.53\linewidth}
    \centering
    \includegraphics[width=\linewidth]{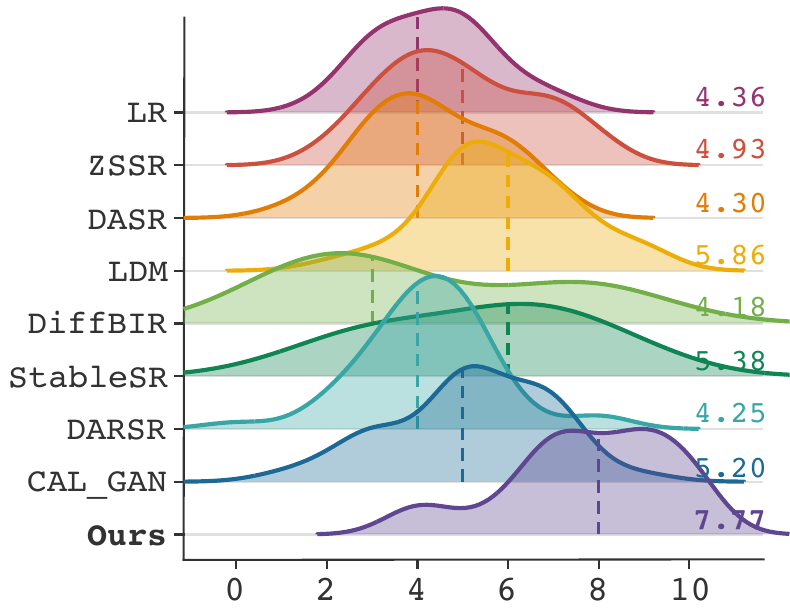} 
    \vspace{-5mm}    
    \caption{User study on the visual perceptual quality of results from different methods on real images.}
    \label{fig:userstudy}
  \end{minipage}
  \hfill 
  \begin{minipage}[t]{0.42\linewidth}
    \centering
    \includegraphics[width=\linewidth]{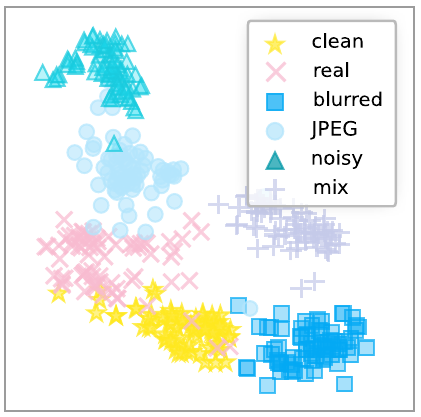} 
    \vspace{-5mm}    
    \caption{t-SNE visualization of embeddings from LR degradation encoder.}
    \label{fig:tsne}
  \end{minipage}
  \vspace{-1mm}
\end{figure}

\noindent \textbf{Training data of LR reconstruction.}
In this section, we demonstrate the robustness of the LR reconstruction network trained with limited data, which forms the cornerstone of our design. As depicted in \tablename~\ref{tab:ablation1}, we incorporated two types of training data. The first category includes synthetic data created using BSRGAN degradation, while the second involves real paired images collected for training. Both settings result in improved performance. 
Specifically, compared to synthetic data, which brings a 0.0486 improvement in LPIPS, the utilization of only 600 images brings a 0.0299 improvement, and 4000 images notably boosts LPIPS by 0.058.
Adding more images beyond this threshold did not yield any further advancement.
We attribute this to the inherent ease in mapping from HR to LR compared to the reverse LR to HR mapping, mitigating the necessity for extensive training data. This assertion finds further support in \figurename~\ref{fig:tsne}, where t-SNE visualization distinctly separates distinct degradations, even for unseen degradation types.

\noindent \textbf{Degradation embedding dimensions.}
\tablename~\ref{tab:dimension} tests different embedding dimensions, indicating that all variants significantly enhance performance. 
While a dimension of 512 (default) is effective, higher one (2048) can further improve results.
%

\noindent \textbf{Our method versus supervised fine-tuning.}
%
To comprehensively illustrate the efficacy of our method, we conduct additional supervised fine-tuning of the baseline model using the gathered real paired data. As depicted in \tablename~\ref{tab:ablationdata}, we note marginal improvements. This aligns with our contention that LR to HR mapping poses inherent difficulties. Training with data from one sensor type showed negligible benefits for another, suggesting a significant gap in degradation patterns. 
This was further corroborated by \figurename~\ref{fig:abaltionvisual}, where it generates over-smoothed outputs. Conversely, our method showcases robustness and substantially enhances the final SR quality.
This proves that our proposed training strategy is more effective.

\begin{figure}[tbp]
  \centering
  \includegraphics[width=1\linewidth]{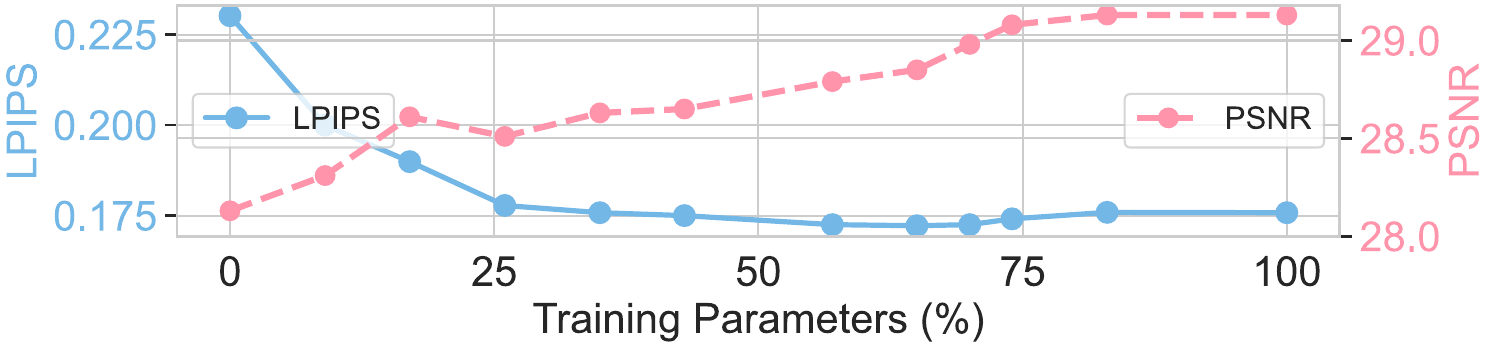} 
  \vspace{-7mm}
  \caption{The performance curve for fine-tuning different percentages of parameters. 
  }
  \label{fig:parameter}
\vspace{-2mm}
\end{figure}

\noindent \textbf{Number of images used in fine-tuning.}
%
We employ self-supervised LR reconstruction fine-tuning on test images to optimize the SR model. This section investigates the impact of the number of fine-tuning images on the final performance. As indicated in \tablename~\ref{tab:ablation2}, we establish a baseline without fine-tuning using ten real-world images. Conducting single-shot fine-tuning on individual images yields the most favorable results, allowing models to best adapt to the distribution of input images. Next, we conduct experiments involving collective fine-tuning of ten images. Results show significant improvements compared to the baseline but are not as effective as fine-tuning individual images separately. Furthermore, we extend our study by fine-tuning the model using an additional forty images to investigate whether acquiring more images from the same sensor would refine the model further. Our findings indicate that compared to training on ten images, there is a decline in LPIPS, while DISTIS and MAD exhibit slight improvements. This suggests a trade-off between fine-tuning performance and efficiency.

\noindent \textbf{High-frequency loss.}
%
\tablename~\ref{tab:ablationloss} illustrates the impact of the introduced high-frequency loss. 
The integration of the high-frequency loss results in a notable improvement, affirming the efficacy of our design. Importantly, it enhances high-frequency recovery and avoids the negative impact of our training method on low-frequency areas.

\noindent \textbf{Fine-tuning parameters.}
%
%
In our exploration of parameter fine-tuning, we observe that increasing the number of trained parameters results in higher PSNR values. However, the LPIPS score reaches its optimal point at approximately 60\% - 70\% of the parameters, as depicted in \figurename~\ref{fig:parameter}. 
Considering the limitation of PSNR, we prioritize the use of LPIPS as our reference. 
It's important to note that different network and testsets may yield varied conclusions. The supplementary materials show more details.

\vspace{-2mm}
\section{Conclusion}
\vspace{-1mm}

In conclusion, our proposed super-resolution training strategy, termed ``Low-Res Leads the Way'', represents an innovative approach that successfully bridges the disparity between synthetic data supervised training and real-world test image self-supervision. Demonstrating impressive performance and robustness across various SR frameworks and real-world benchmarks, our method marks a significant advancement toward achieving effective real-world applications.

{
    \small
    \bibliographystyle{ieeenat_fullname}
    \bibliography{main}
}

\clearpage
\setcounter{page}{1}
\setcounter{section}{0} 
\renewcommand\thesection{\Alph{section}}
\renewcommand\thesubsection{\thesection.\arabic{subsection}}

\maketitlesupplementary

\section{Experimental Details}

\subsection{Fine-tuning Details}

\noindent Due to the different network architectures of different types of models, we trained different parts of the parameters for them.
The rationale behind parameter selection for training is corroborated by empirical experiments detailed further in the text.

\begin{itemize}
    \item For the training phases specific to BSRGAN, Real-ESRGAN+, and SwinIR-GAN, selective freezing of initial layers is implemented to concentrate training on the deeper parameters.
    \item In the case of FeMaSR, which is based on the VQGAN (Vector Quantized Generative Adversarial Network) structure, the focus is placed on the parameters of the VQGAN encoder.
    \item StableSR, which utilizes a pre-trained diffusion model, applies a controllable feature wrapping (CFW) module with an adjustable coefficient to refine the outputs of the diffusion model during the decoding process of the autoencoder. We choose to fine-tune the designed Collaborative Feature Weighting module and part of the encoder.
\end{itemize}

It usually takes 150 to 500 iterations to train. The time depends on the baseline network size, ranging from seconds to a few minutes. 
Our method can be fine-tuned either on individual images or on the entire test set assuming consistent degradation across the test set, which greatly reduces computational cost.
\tablename~\ref{tab:runtime} shows that our method takes only 8 minutes to fine-tune on the whole test set, much faster than others. Individual fine-tuning can improve the results if needed.

\subsection{Testing Datasets}

\noindent The validation of the effectiveness of our training method in real-world scenarios is conducted using real-world paired datasets, RealSR~\cite{cai2019toward}, and DRealSR~\cite{wei2020component}. These datasets are meticulously curated from various sensors to reflect different degradation characteristics inherent in each device. 
Furthermore, the datasets are segmented based on the capturing equipment. For RealSR, a 2$\times$ scale factor is employed, with separate subsets for Canon and Nikon. In the case of DRealSR, a 4$\times$ scale is applied across three subsets corresponding to Sony, Panasonic, and Olympus.
To ensure a fair comparison with other models, we follow common settings employed by most methods. Each image is segmented into multiple smaller patches for performing 4$\times$ super-resolution, with the patch size for LR images being 128$\times$128 and for HR images being 512$\times$512.

\begin{table}
    \centering
    \resizebox{0.8\linewidth}{!}{%
    \begin{tabular}{@{}lcccc@{}}
        \toprule
                         & LPIPS{\color[HTML]{369DA2} $\downarrow$}  
                         & DISTS{\color[HTML]{369DA2} $\downarrow$}   
                         & PSNR{\color[HTML]{369DA2} $\uparrow$}  
                         & \textbf{fine-tuning time}{\color[HTML]{369DA2} $\downarrow$} 
                         \\ \midrule
        Ours ($d=2048$)      & \textbf{0.1629} & \textbf{0.1630} & \textbf{29.56} & \textbf{8 min} \\
        ZSSR                 & 0.2424 & 0.1889 & 29.14 & 19 \textcolor{t}{hr} \\
        KernelGAN+ZSSR       & 0.3315 & 0.2774 & 23.52 & 72 \textcolor{t}{hr} \\ 
        deep plug-and-play   & 0.2604 & 0.2524 & 29.44 & 1.4 \textcolor{t}{hr} \\
        deep image prior     & 0.2091 & 0.2054 & 29.32 & 28 \textcolor{t}{hr} \\
        \bottomrule
        \end{tabular}%
    }
    \caption{Our method can be fine-tuned on the entire test set assuming consistent degradation across the test set, which greatly reduces computational cost. Other methods need to be trained on each individual image.}
    \label{tab:runtime}
\end{table}

\subsection{Evaluation Metrics}

\noindent Following prior research~\cite{jinjin2020pipal, gu2020image}, our study adopts a carefully curated set of perceptual metrics, ones that have shown a higher correlation with human perception, including Learned Perceptual Image Patch Similarity (LPIPS)~\cite{zhang2018unreasonable}, Deep Image Structure and Texture Similarity (DISTS)~\cite{ding2020image}, and Normalized Laplacian Pyramid Distance (NLPD)~\cite{hepburn2019enforcing}. 
LPIPS and DISTS have been empirically validated in~\cite{jinjin2020pipal, gu2020image} as more closely aligned with human visual assessment than other metrics. We also include traditional metrics, such as Peak Signal-to-Noise Ratio (PSNR), Structural Similarity Index (SSIM)~\cite{wang2004image}, and most apparent distortion (MAD)~\cite{larson2010most}.

A previous study \cite{jinjin2020pipal, gu2020image} investigated the correlation between human visual perception quality of images and various Image Quality Assessment (IQA) metrics, with the findings summarized in the \tablename~\ref{tab:iqa}. The experimental results reveal that MAD, LPIPS, and DISTS outperform traditional PSNR and SSIM across various aspects in the context of super-resolution evaluation. Specifically, MAD demonstrates superior accuracy in assessing traditional SR methods. On the other hand, both LPIPS and DISTS exhibit higher precision when evaluating GAN-based SR methods. In the overall comparison, DISTS emerges as the most effective metric for super-resolution assessment. These findings underscore the limitations of relying solely on conventional metrics such as PSNR and SSIM, emphasizing the importance of incorporating newer metrics like MAD, LPIPS, and DISTS for a more comprehensive and accurate evaluation of super-resolution techniques.

\begin{table}
    \centering
    \resizebox{\linewidth}{!}{%
    \begin{tabular}{lcccc}
    \toprule
      Method  & SR Full & Traditional SR & PSNR. SR & GAN-based SR\\ \midrule \midrule
      PSNR        & 0.4099 & 0.4782 & 0.5462 & 0.2839\\
      SSIM        & 0.5209 & 0.5856 & 0.6897 & 0.3388\\
      MAD         & 0.5424 & \textbf{0.6720} & 0.7575 & 0.3494\\ \midrule
      LPIPS       & 0.5614 & 0.5487 & 0.6782 & 0.4882\\
      DISTS       & \textbf{0.6544} & 0.6685 & \textbf{0.7733} & \textbf{0.5527}\\ \bottomrule
    \end{tabular}
    }
    \caption{The Spearman rank correlation coefficient (SRCC) between MOS (Mean Opinion Score) and various IQA (Image Quality Assessment) metrics across different distortion sub-types.}
    \label{tab:iqa}
\end{table}

\section{LR Reconstruction Network}

\subsection{Degradation Encoder}

Following the methodology proposed by Liu et al.~\cite{liu2023degae}, our degradation encoder is constructed by integrating a pre-trained SR-GAN model~\cite{ledig2017photo} and downsampling layers. This collaborative framework aims to produce degradation embeddings, denoted as $e$, with a dimensionality of 512. The choice of a relatively small dimension for $e$ ensures that the degradation embeddings do not encapsulate intrinsic image information but are sufficiently representative of content pertaining specifically to the degradation process. This design principle is crucial in isolating and preserving only the features relevant to degradation, avoiding contamination with the original image characteristics.

\subsection{Reconstructor}

In our methodology, we incorporate a modulation-demodulation-convolution strategy reminiscent of Instance Normalization as employed in StyleGAN2~\cite{karras2020analyzing}. This approach effectively utilizes the degradation embedding $e$ to facilitate LR reconstruction when combined with the SR network's output $I_{SR}$.
To delve deeper into the specifics of this strategy, during modulation, a style is learned from the provided degradation embedding $e$. The modulation operation scales each input feature map of the convolution using the acquired style, as denoted by the equation 
$$
w_{i j k} = s_i \cdot w_{i j k},
$$ 
the variables $w$ and $w^{\prime}$ represent the original and modulated weights, respectively. 
The scale factor, denoted as $s_i$, corresponds to the $i$th input feature map. The indices $j$ and $k$ are used to iterate over the output feature maps and spatial footprint of the convolution, respectively. 
This modulation process ensures that the convolutional features are adaptively adjusted based on the characteristics embedded in the degradation embedding.
Following modulation, a demodulation step is executed to obtain the demodulated convolution weights, represented as 
$$
w_{i j k}^{\prime \prime} = \frac{w_{i j k}^{\prime}}{\sqrt{\sum_{i, k} w_{i j k}^{\prime}{ }^2+\epsilon}}.
$$
The primary objective of demodulation is to restore the outputs to a unit standard deviation, providing stability and normalizing the feature representations.
It is crucial to emphasize that this modulation-demodulation-convolution strategy facilitates the integration of degradation-specific information into the LR reconstruction process. The adaptability of the convolutional features based on the learned style ensures that the network can effectively reconstruct LR inputs, enhancing the overall performance of the SR framework.

\begin{figure}[tbp]
  \centering
  \includegraphics[width=0.9\linewidth]{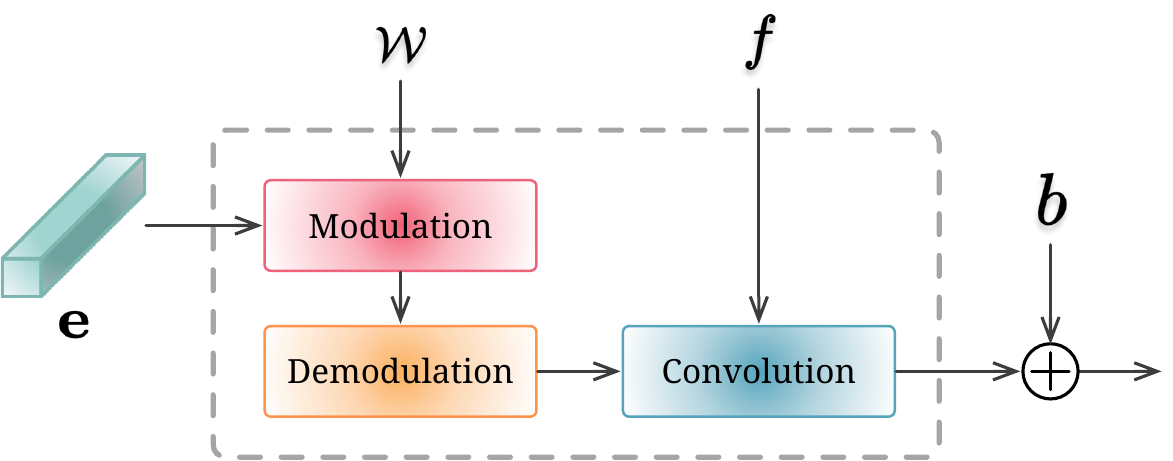} 
  \caption{Modulation method used in our LR reconstructor.}
  \label{fig:mod}
\end{figure}

\section{More Experiment Results}

\subsection{The Effect of Fine-tuning Parameters for Different Network Architecture.
}
\begin{figure}[tbp]
  \centering
  \includegraphics[width=0.9\linewidth]{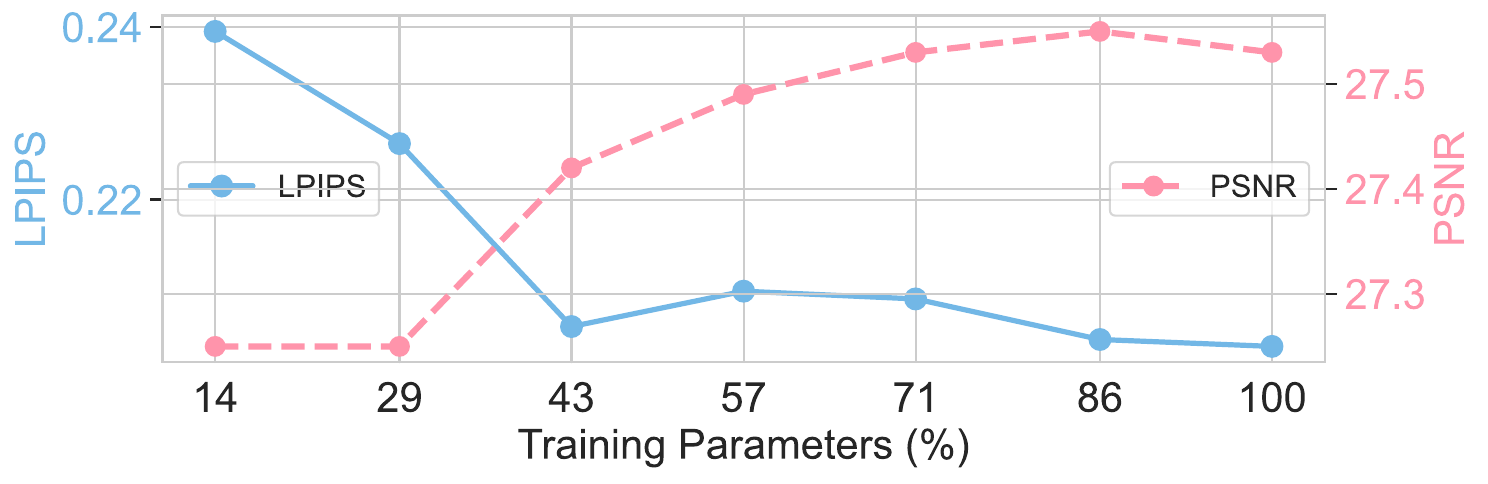} 
  \caption{The performance curve for fine-tuning different per-
centages of parameters for FeMaSR.}
  \label{fig:plotfemasr}
\end{figure}

\begin{figure}[tbp]
  \centering
  \includegraphics[width=0.9\linewidth]{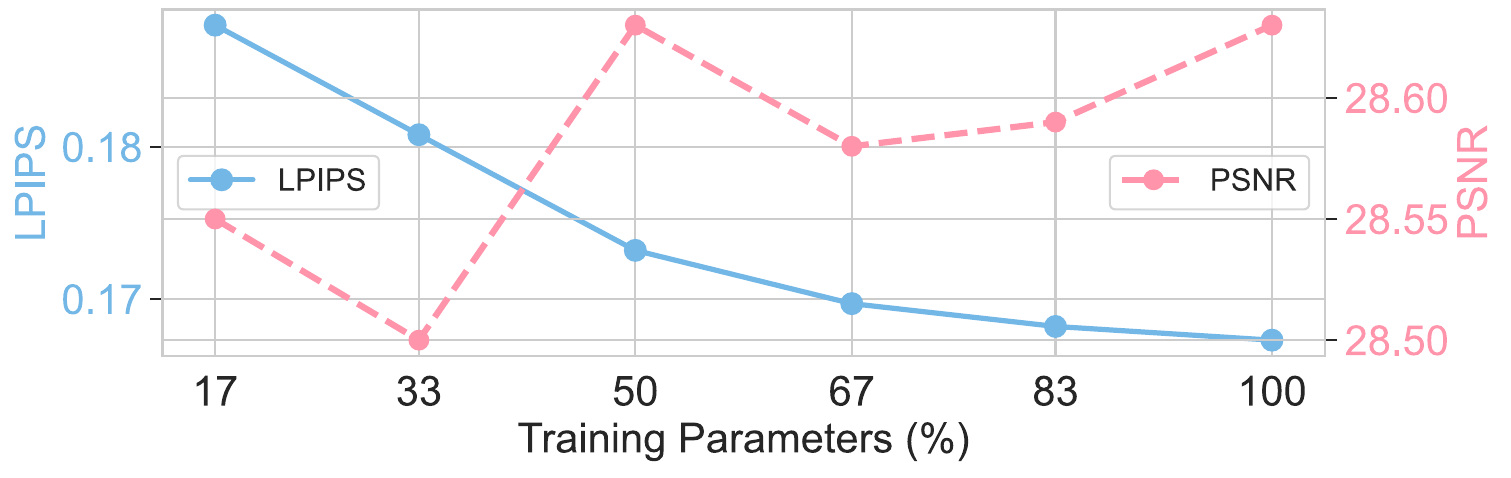} 
  \caption{The performance curve for fine-tuning different per-
centages of parameters for SwinIR.}
  \label{fig:plotswinir}
\end{figure}

In our investigation into the impact of training parameters on the performance of the FeMaSR and SwinIR networks, the influence is shown in \figurename~\ref{fig:plotfemasr} and \figurename~\ref{fig:plotswinir}. Specifically, for the FeMaSR network, the optimal PSNR is achieved when training parameters constitute 86\%, while the optimal LPIPS is obtained at 100\%. In contrast, SwinIR attains the best PSNR and LPIPS values almost simultaneously at 100\% of training parameters.

\subsection{Ablation on Model Size} 
Table~\ref{tab:size} delineates the efficacy of our proposed model across a spectrum of sizes, demonstrating that our method retains robust performance notwithstanding the model's capacity. From a comprehensive model with 12.9 million parameters to a compact version with merely 495 thousand parameters, our approach consistently outperforms the baseline.

\begin{table}
    \centering
    \resizebox{\linewidth}{!}{%
    \begin{tabular}{@{}lcccccc@{}}
    \toprule
                    & Paras  & FLOPS   
                    & PSNR{\color[HTML]{369DA2} $\uparrow$}  
                    & MAD{\color[HTML]{369DA2} $\downarrow$}    
                    & LPIPS{\color[HTML]{369DA2} $\downarrow$}  
                    & DISTS{\color[HTML]{369DA2} $\downarrow$}  \\ \midrule
    baseline        & -      & -       & 28.13 & 118.48 & 0.2302 & 0.2102 \\
    + LWay \texttt{(Large)}    & 12.9 M & 589.4 G & 28.85 & 104.71 & 0.1722 & 0.1772 \\
    + LWay \texttt{(Medium)}   & 5.38 M & 117.6 G & 29.50 & 99.76  & 0.1798 & 0.1810 \\
    + LWay \texttt{(Small)}    & 2.77 M & 44.49 G & 28.69 & 106.42 & 0.1837 & 0.1862 \\
    + LWay \texttt{(Tiny)}     & 495 K  & 19.38 G & 28.70 & 104.92 & 0.1808 & 0.1842 \\ \bottomrule
    \end{tabular}%
    }
    \caption{The performence of different model size. LWay is not contingent on the parameter count of the LR reconstruction, demonstrating effectiveness even with a small parameter volume.}
    \label{tab:size}
\end{table}

\subsection{Performance on Different Degradation} 
Table~\ref{tab:moredegradation} demonstrates the robustness of our 'LWay' method in handling various types of image degradations. It presents notable improvements in PSNR and reductions in LPIPS for both real-world degradation and synthetic distortions such as blurring and JPEG compression, signifying our method's efficacy in maintaining image integrity across different degradation scenarios.

\begin{table}[]
\centering
\resizebox{\linewidth}{!}{%
\renewcommand{\arraystretch}{1.25}
\begin{tabular}{l|cc|cc|cc|cc}
\hline
\multirow{2}{*}{}         & \multicolumn{2}{c|}{real-world degradation} & \multicolumn{2}{c|}{\textbf{synthetic}, blur 17$\times$17} & \multicolumn{2}{c|}{\textbf{synthetic}, blur 11$\times$11} & \multicolumn{2}{c}{\textbf{synthetic}, JPEG $q=15$} \\ \cline{2-9} 
                          & PSNR                 & LPIPS                & PSNR                        & LPIPS                        & PSNR                        & LPIPS                        & PSNR                     & LPIPS                    \\ \hline
baseline                  & 28.13                & 0.2302               & 27.55                       & 0.4065                       & 27.5                        & 0.3922                       & 26.60                    & 0.4240                   \\
\textbf{+LWay} ($d$=2048) & 29.56                & 0.1629               & 28.39                       & 0.2755                       & 29.02                       & 0.2265                       & 26.85                    & 0.3122                   \\ \hline
\end{tabular}%
}
\caption{Performance on different degradation.  LWay improve image quality under a range of degradations.}
\label{tab:moredegradation}
\end{table}



\section{More Visual Results}

\subsection{LR Reconstruction Visualization}

\begin{figure*}[tbp]
  \centering
  \vspace{5mm}
  \includegraphics[width=1\linewidth]{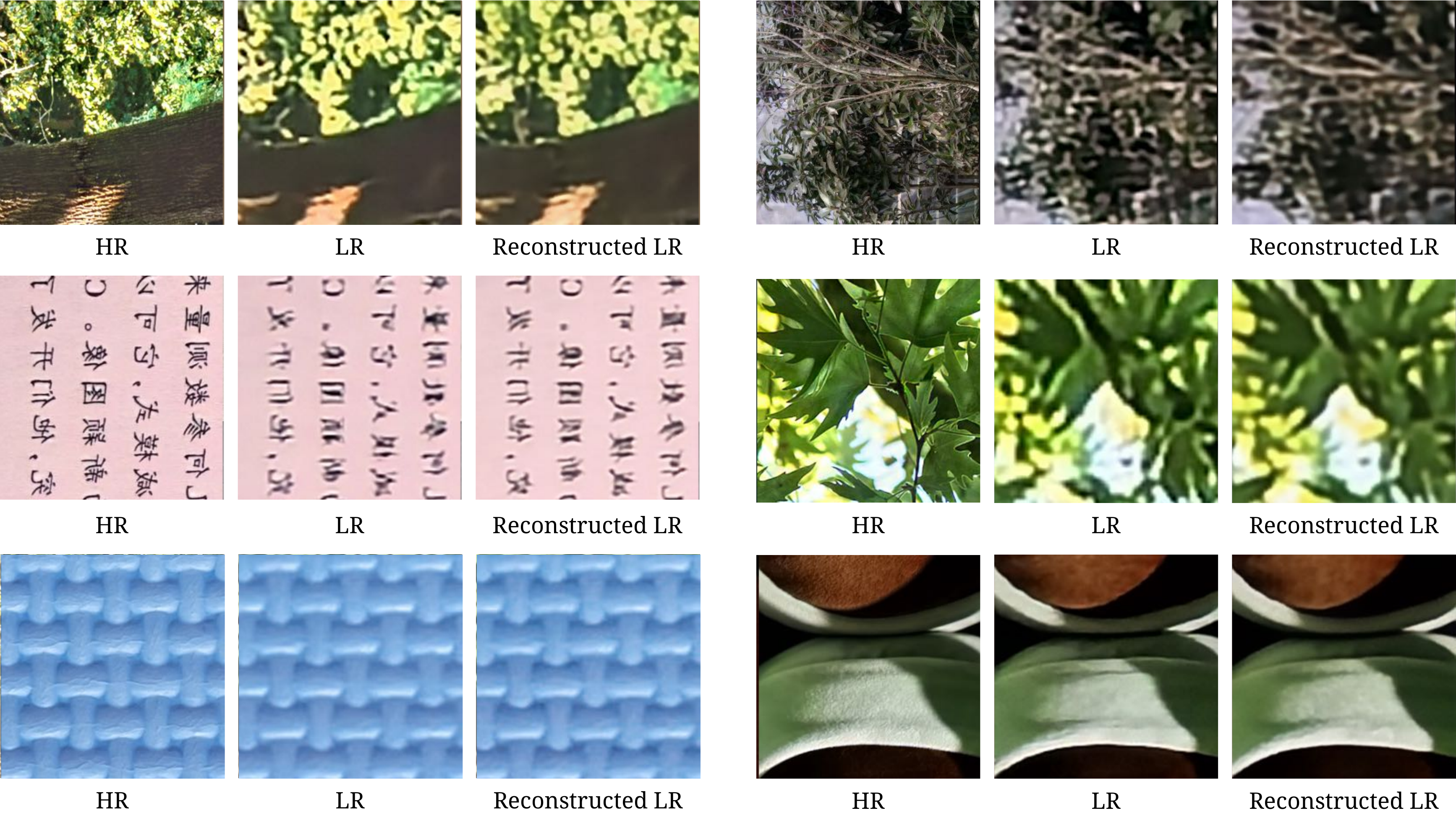} 
  \vspace{3mm}
  \caption{Visual results of LR reconstruction. For instance, given an input LR image and HR image, the degradation encoder encodes a 512-dimension degradation embedding $e$, the reconstructor utilizes $e$ and the HR image to reconstruct the estimated LR image.}
  \label{fig:lr}
  \vspace{3mm}
\end{figure*}

The visual outcomes of the LR reconstruction network are illustrated in \figurename~\ref{fig:lr}, encompassing HR, LR, and the reconstructed LR images. Notably, our network demonstrates the capability to restore LR images that closely approximate the ground truth LR by extracting a 512-dimensional degradation embedding solely from the LR input and subsequently integrating it with the HR image. This process demonstrates the effectiveness of our LR reconstruction approach in achieving visually compelling results.
The showcased robustness of our LR reconstruction network is particularly noteworthy. Given that the transition from HR to LR is generally considered easier compared to the reverse process, our method exhibits a heightened degree of resilience with limited data. Leveraging only a finite dataset, our approach achieves a robust performance, underscoring its capacity to generalize and adapt well to diverse LR input scenarios.

\subsection{More Visual Comparison}

\begin{figure*}[tbp]
  \centering
  \includegraphics[width=0.92\linewidth]{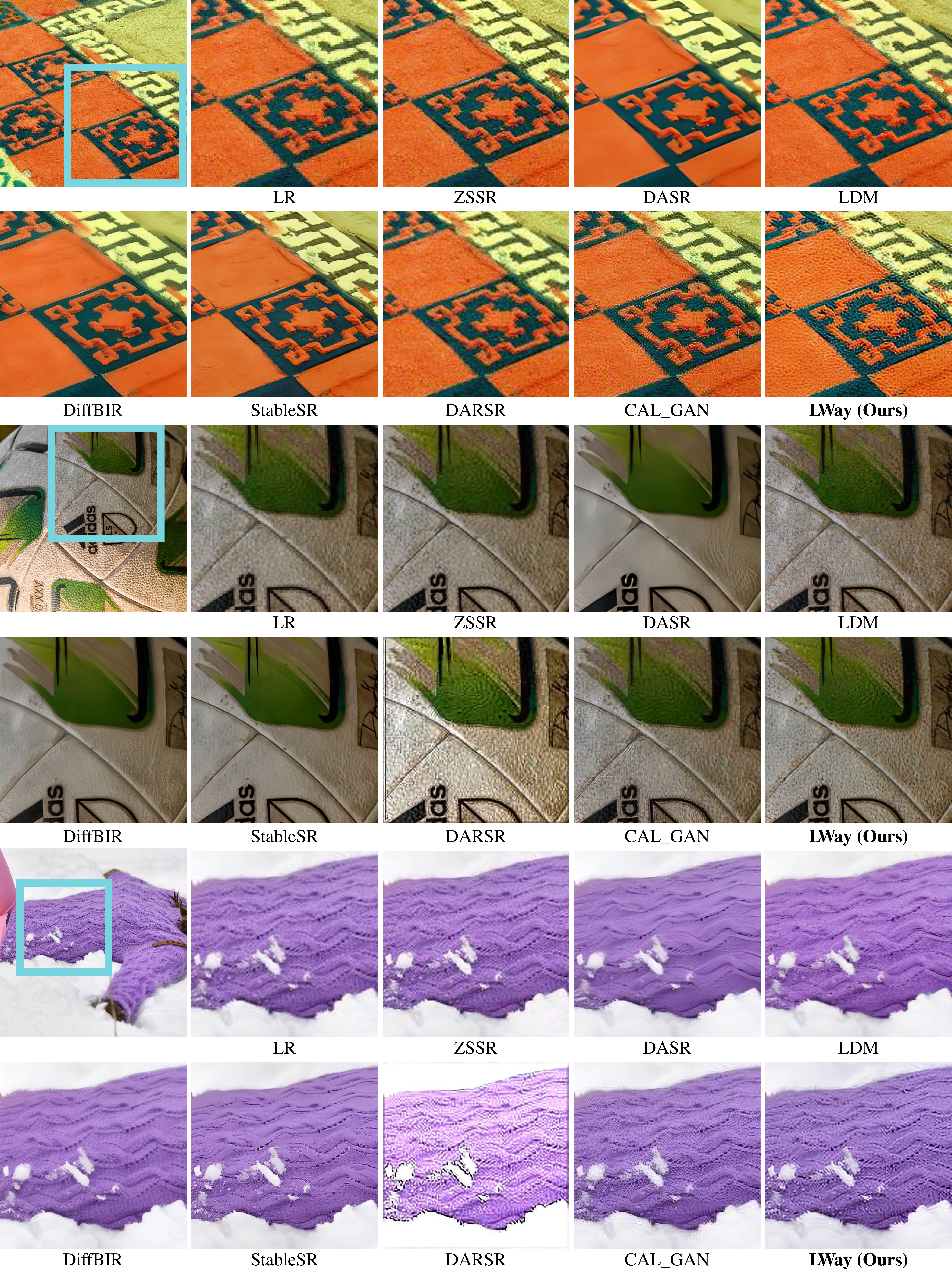} 
  \caption{More visual comparisons.}
  \label{fig:visualsupp1}
\end{figure*}

\begin{figure*}[tbp]
  \centering
  \includegraphics[width=0.92\linewidth]{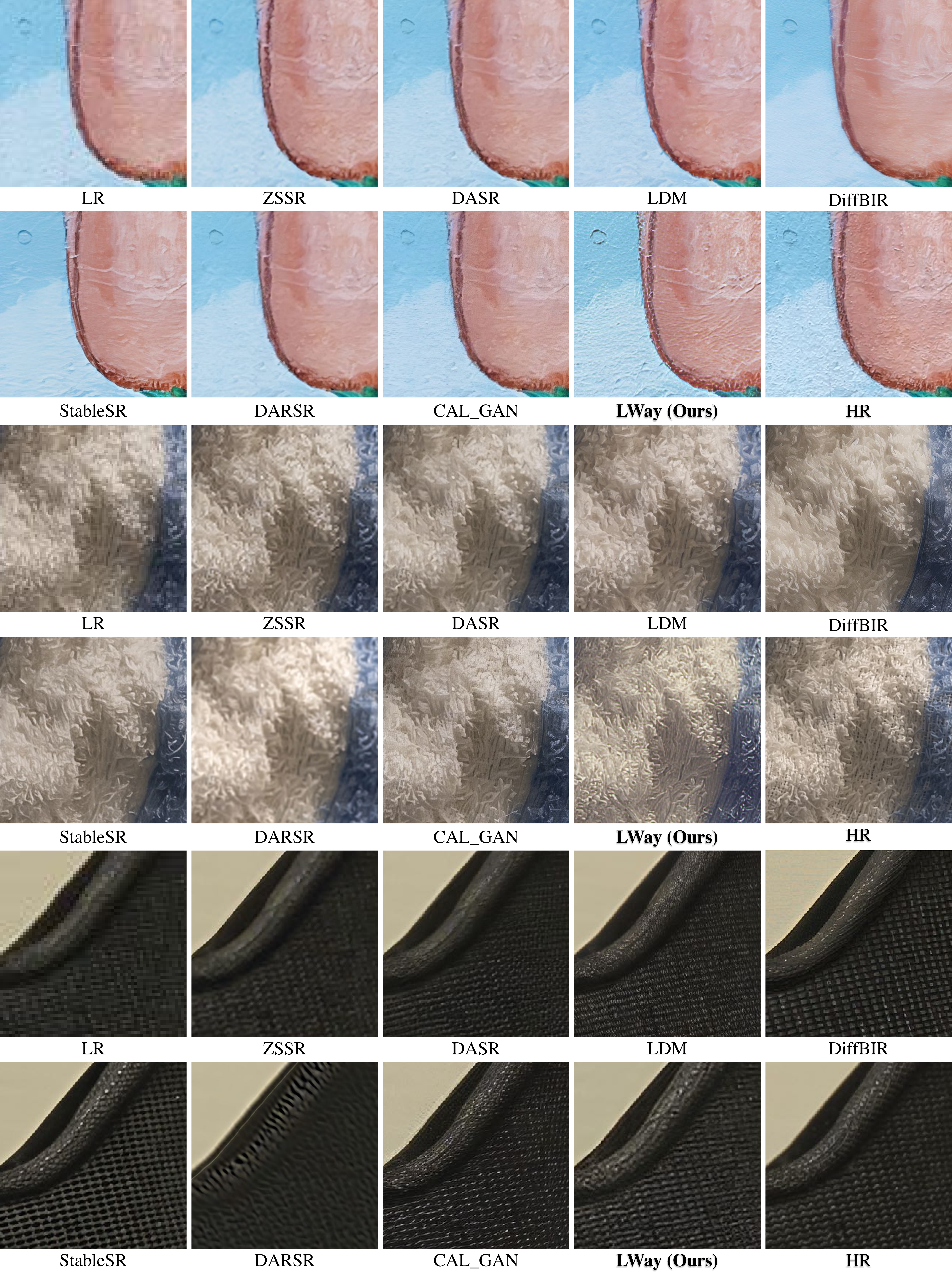} 
  \caption{More visual comparisons.}
  \label{fig:visualsupp2}
\end{figure*}

\figurename~\ref{fig:visualsupp1} and \figurename~\ref{fig:visualsupp2} provide additional comparisons of our proposed method with other state-of-the-art (sota) approaches. Our method excels in effectively restoring the texture and fine details of images. In contrast, DASR, DiffBIR, and StableSR tend to produce smoother results at the expense of losing texture details. ZSSR, on the other hand, exhibits limited restoration capabilities, resulting in less clear outcomes that are less faithful to the LR input. The results generated by LDM display inconsistencies in texture details compared to the ground truth. DARSR, while prone to failure and introducing significant color bias, and CAL\_GAN both exhibit varying degrees of artifacts in their outputs.
These visual comparisons underscore the superior performance of our proposed method in preserving intricate details and textures during the super-resolution process. The tendency of other methods to sacrifice fine details for smoother results, introduce artifacts, or inaccurately represent texture details highlights the unique strengths of our approach. 

\section{Discussion}

\noindent \textbf{Differences to optimization-based methods.}
Optimization-based methods, relying on pre-defined degradation models or downsampling operators, have limited capabilities in handling complex degradation. 
They are also time-consuming and difficult with large data.
In contrast, our approach incorporates a more general and robust degradation modeling. Moreover, our method marries the benefits of supervised and unsupervised training, outperforming optimization-based methods that only use test images.

\noindent \textbf{Differences to KernelGAN.}
KernelGAN’s discriminator only makes binary judgments (0/1), while LWay uses pixel-level regression to better capture the distribution. Moreover, local KernelGAN’s kernels have limited information and robustness in real-world, while our embedding has richer external priors rather than relying on solely learning test images and is robust as demonstrated by validation.

\section{Limitation}

The proposed architecture excels in extracting and restoring information from low-resolution (LR) images, especially when they contain discernible texture details. It is within these conditions that our method showcases its maximum effectiveness. However, a limitation might arise when the LR images themselves lack texture details, impeding the model's capability to execute effective restoration.

\end{document}